\documentclass[12pt,preprint]{aastex}
\usepackage{color}

\begin{document}

\title{Discovery of A Very Bright, Strongly-Lensed $z=2$ Galaxy in 
       the SDSS DR5}


\author{
Huan Lin\altaffilmark{1},
Elizabeth Buckley-Geer\altaffilmark{1},
Sahar S.\ Allam\altaffilmark{1,2},
Douglas L.\ Tucker\altaffilmark{1},
H.\ Thomas Diehl\altaffilmark{1},
Donna Kubik\altaffilmark{1},
Jeffrey M.\ Kubo\altaffilmark{1},
James Annis\altaffilmark{1},
Joshua A.\ Frieman\altaffilmark{1,3},
Masamune Oguri\altaffilmark{4},
Naohisa Inada\altaffilmark{5}}





\altaffiltext{1}{Fermi National Accelerator Laboratory,   P.O. Box 500, Batavia, IL 60510}
\altaffiltext{2}{University of Wyoming, Dept.\ of Physics \& Astronomy, P.O.Box 3905, Laramie, WY 82071}
\altaffiltext{3}{Kavli Institute for Cosmological Physics and
Department of Astronomy and Astrophysics, University of Chicago, 5640 South Ellis Avenue, Chicago, IL 60637}
\altaffiltext{4}{Kavli Institute for Particle Astrophysics and
Cosmology, Stanford University, 2575 Sand Hill Road, Menlo Park, CA 94025}
\altaffiltext{5}{Cosmic Radiation Laboratory, RIKEN, 2-1 Hirosawa, Wako, 
Saitama 351-0198, Japan}


\begin{abstract}

We report on the discovery of a very bright $z = 2.00$ 
star-forming galaxy that is strongly lensed by a foreground $z=0.422$
luminous red galaxy (LRG).  This system was found in a systematic search for 
bright arcs lensed by LRGs and brightest cluster galaxies
in the Sloan Digital Sky Survey Data Release 5 sample.
Follow-up observations on the Subaru 8.2m telescope on Mauna Kea 
and the Astrophysical Research Consortium 3.5m telescope at 
Apache Point Observatory confirmed the lensing nature of this system. 
A simple lens model for the system, assuming a singular isothermal 
ellipsoid mass distribution, yields an Einstein radius of
$\theta_{\rm  Ein}=3.82\pm{0.03} \arcsec$ or $14.8\pm{0.1}h^{-1}$~kpc 
at the lens redshift.  The total projected mass enclosed within the
Einstein radius is $2.10\pm{0.03}\times10^{12} h^{-1}  M_{\sun}$, and
the magnification factor for the source galaxy is $27 \pm 1$.
Combining the lens model with our $gVriz$ photometry, we find an
(unlensed) star formation rate for the source galaxy
of $32 \ h^{-1} \ M_\sun \ {\rm yr}^{-1}$, adopting a 
fiducial constant star formation rate model with an age of 
100 Myr and $E(B-V) = 0.25$.  With an apparent magnitude of $r = 19.9$,
this system is among the very brightest lensed $z \geq 2$ galaxies,
and provides an excellent opportunity to pursue detailed studies of the 
physical properties of an individual high-redshift star-forming galaxy.

\end{abstract}





\section{Introduction}

Strong lensing systems provide the dual opportunity to study
both the foreground mass distribution along the line of sight to the lens
and the physical properties of the background object 
that is being lensed.  The latter is especially
useful in studies of high-redshift galaxies, for which lensing
provides a vital boost in the apparent brightness of these
faint objects, which are otherwise difficult to study in detail.

For many years the $z = 2.72$ system cB58 \citep{yee96} served
as the prototypical lensed high-redshift Lyman break galaxy 
\citep[LBG; e.g.,][]{steidel03}.  At $r = 20.4$, it is very bright
and thereby allowed a number of detailed studies of the physical
properties of a {\em single} LBG to be carried out 
\citep[e.g.,][]{pettini00,teplitz00}.  Recently a number
of high-redshift lensed systems have been discovered, either serendipitously 
or in systematic searches, that are brighter than cB58
\citep{smail07,belokurov07}, including the current record holder,
the ``8~o'clock arc,'' at $r = 19.2$ \citep{allam07}.
These discoveries have been enabled by the
Sloan Digital Sky Survey \citep[SDSS;][]{york00}, 
which provides the very large search area needed to systematically find 
these rare examples of extremely bright lensed high-redshift galaxies.



In this paper we report on the discovery of another remarkably bright
($r = 19.9$) strongly lensed $z = 2.00$ galaxy, the first system
we have confirmed from a systematic search program for very bright lensed 
arcs that we are carrying out using the SDSS data.
This paper is organized as follows: \S\ref{sec:search} describes
the arc search and the discovery, \S\ref{sec:follow_up} describes the 
follow-up imaging and spectroscopy that led to confirmation of the system as a
gravitational lens, \S\ref{sec:modeling} describes the modeling of
the system including the photometry measurements, \S~\ref{sec:SFR} 
describes the source galaxy star formation rate measurements, and
finally \S\ref{sec:conclusions} presents our conclusions.
We assume a flat cosmology with $\Omega_M =  0.3$, 
$\Omega_{\Lambda}=0.7$,   and $H_0=100h$~km~s$^{-1}$~Mpc$^{-1}$, 
unless otherwise noted.

\section{Arc Search Sample}\label{sec:search}
The  SDSS \citep{york00}  is a  digital imaging and spectroscopic
survey that,  over the  course of five  years, mapped nearly  one
quarter of  the   celestial  sphere  in  five  filter  bands 
\citep[$ugriz$;][]{fukugita96} down to $r=22.2$ and obtained spectra for
$\approx10^6$  astronomical objects \citep{dr5}.
The SDSS completed its first  phase of operations in June 2005
and recently completed a three-year extension known as SDSS-II 
in July 2008. (For more details, please consult {\tt www.sdss.org}.)

We previously reported the serendipitous discovery in the SDSS 
Data Release 4 \citep[DR4;][]{dr4} of the brightest lensed Lyman break 
galaxy (LBG) currently known, the 8 o'clock arc \citep{allam07}. 
The LBG in that system is at a redshift of 2.73 and is lensed by a luminous 
red galaxy (LRG) at a redshift of 0.38.  
The three bright gravitationally lensed images 
have a total magnitude of $r = 19.2$ and are quite blue
($g-r = 0.7$).  Motivated by this discovery and using the characteristics of 
the 8~o'clock arc system as our starting point, we have conducted a 
systematic search \citep{kubik07}\footnote{\url{http://www.physics.niu.edu/physics/academic/grad/theses/Donna.pdf}}
for similar systems in the 
SDSS Data Release 5 \citep[DR5;][]{dr5}.
The search started from two catalogs:  the first consisting of
221,000 LRGs derived from the SDSS database and the second consisting
of 29,000 brightest cluster galaxies (BCGs) compiled by one of us (J.~Annis)
using an earlier version \citep{hansen05} of the maxBCG cluster finding 
technique \citep{koester07}.  We defined a database query which was run 
on the DR5 Catalog Archive Server (CAS) database. 
This query searched for LRGs and BCGs which have one or more neighboring blue 
objects, defined using color cuts $g-r < 1$ and $r-i < 1$, 
that were detected by the SDSS photometric pipeline within
a search radius of $10\arcsec$.  
We note that due to issues of seeing and object deblending in the SDSS,
our search will effectively find systems with Einstein radii larger
than about $2\arcsec$ or so.  Our search is therefore complementary
to a spectroscopic lensing survey like the Sloan Lens ACS Survey 
\citep[SLACS;][]{bolton06}, 
which is limited to systems with image separations smaller than the $3\arcsec$
SDSS spectroscopic fiber diameter.

\begin{figure}
\plotone{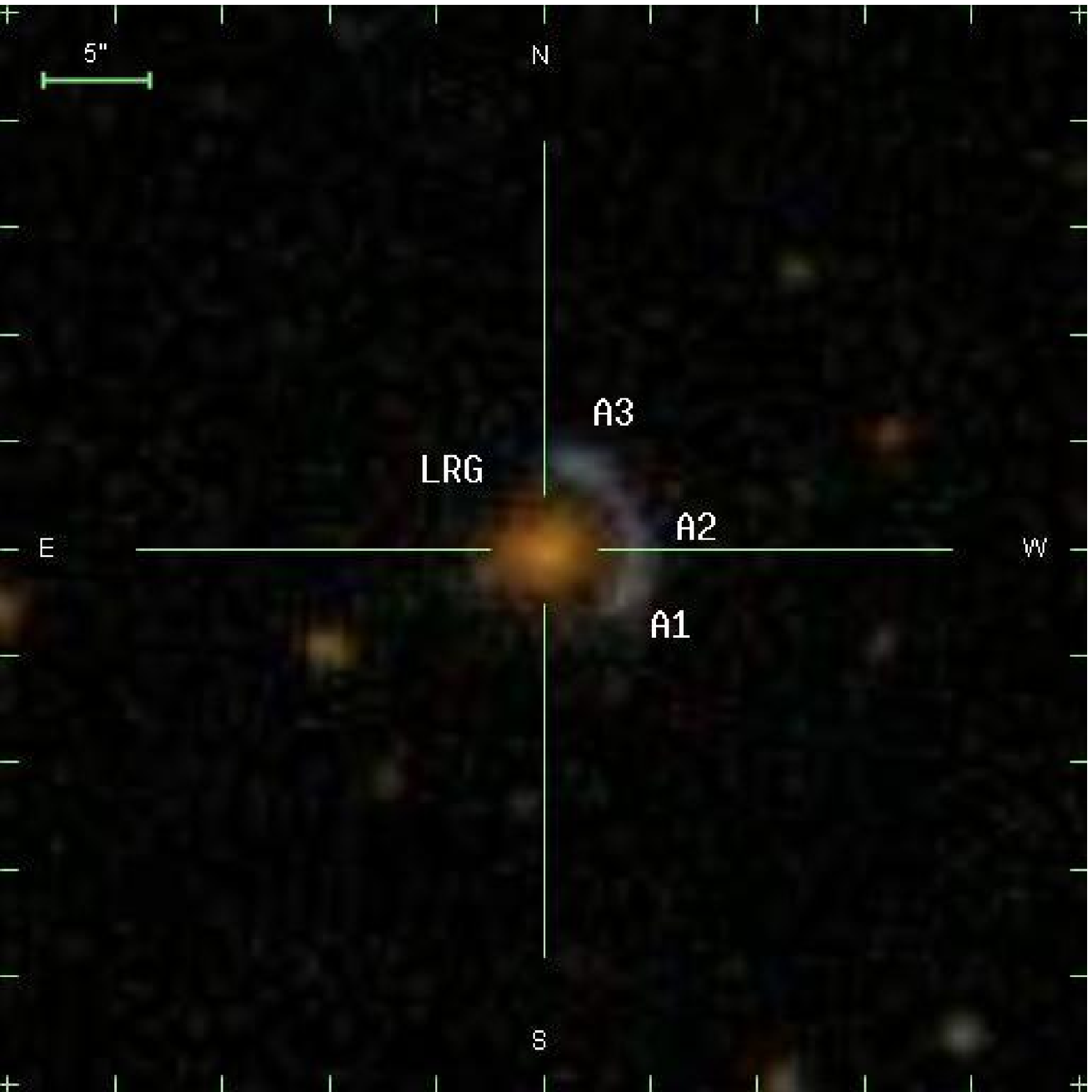}
\caption{The SDSS $gri$ color composite image (provided by the SDSS 
{\tt SkyServer}) from which the arc was discovered.  Labels have been
added to indicate the locations of the three lensed arc components 
(A1, A2, and A3) and the position of the LRG.
\label{sdss_image}}
\end{figure}

\begin{figure}
\plotone{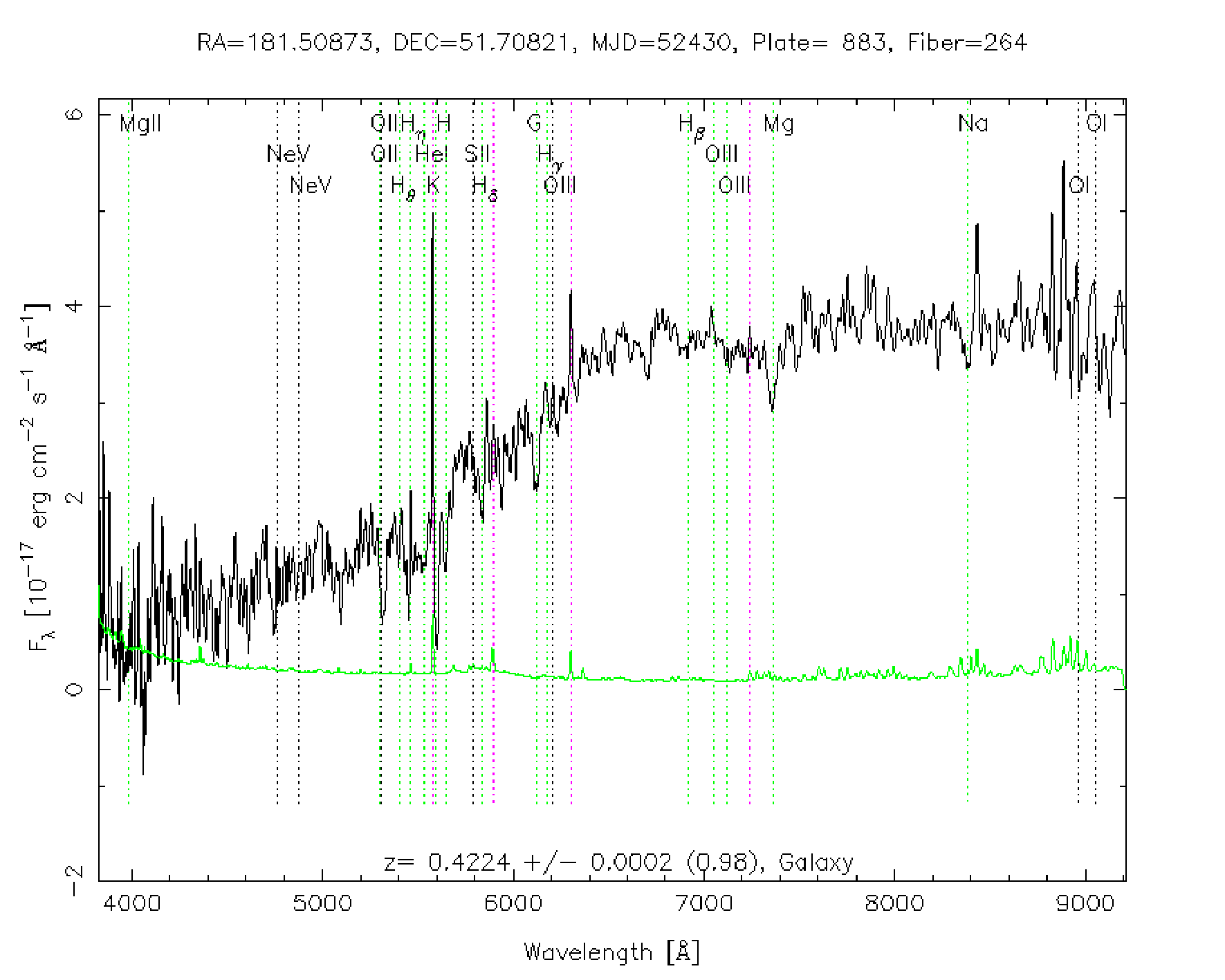}
\caption{The SDSS spectrum (provided by the SDSS {\tt SkyServer})
of the LRG, showing an early-type galaxy spectrum with a redshift 
$z = 0.4224 \pm 0.0002$.  The labels and vertical green lines
indicate potential spectral features (whether present or not)
at the LRG redshift.  Note that the prominent emission feature
at $\sim$5577\AA \ is really a strong night sky line subtraction residual, 
as indicated by the vertical magenta lines.
\label{lrg_spectrum}}
\end{figure}

Our query returned 57,485 systems, which
were then ranked by the number of blue objects, $n$.  The 1081 systems
with $n \ge 3$ were inspected by four separate inspectors who looked
for arc-like morphology in the SDSS CAS $gri$ color jpeg images. 
The 14 final candidates found in this sample have already been described in
\cite{kubik07}, including an initial analysis of their Einstein radii and
mass-to-light ratios.  To date we have spectroscopically 
confirmed 6 of them as lensed, including 3 with source redshifts $z \geq 2$.
Additional details of follow-up observations and 
lens modeling for these systems, as well as for other systems found in
a separate search of a sample of SDSS interacting/merging galaxies,
will be the subject of future papers.
One inspector also examined the 7442 systems 
in the $n = 2$ sample, which yielded the object described in this paper.
This system was the brightest and most striking arc candidate 
from the $n = 2$ list, and we dubbed the system the ``Clone'' 
as it was very similar to the 8~o'clock arc in morphology and 
brightness.  In Fig.~\ref{sdss_image} we
show the discovery SDSS image of this system.  The lensing LRG is
the object SDSS J120602.09+514229.5, and Fig.~\ref{lrg_spectrum}
shows its SDSS spectrum, indicating absorption features of an early type 
elliptical galaxy at a redshift of 0.422. 
The very blue arc knots, labeled  A1 through A3 in Fig.~\ref{sdss_image}, 
are associated with two objects identified by the SDSS photometric pipeline
(A1/A2 = SDSS J120601.69+514227.8 and A3 = SDSS J120601.93+514233.5),
but they are not SDSS spectroscopic targets and so do not have any 
SDSS spectroscopic redshifts.

\section{Follow-up Imaging and Spectroscopy}\label{sec:follow_up}

In order to confirm the Clone system as a gravitational lens we have carried 
out a follow-up program of imaging and
spectroscopy using the Astrophysical Research Consortium (ARC) 3.5m telescope
at Apache Point Observatory (APO) and the 8.2m Subaru telescope on Mauna Kea.

\subsection{Subaru Imaging and Spectroscopy}

Initial follow-up imaging and long-slit spectroscopy were carried out with the 
Faint Object Camera and Spectrograph (FOCAS) instrument on the Subaru 8.2m
telescope \citep{kashikawa02}; see the observation log in 
Table~\ref{table_obslog}.
The instrument has a $6\arcmin$ circular field of view and the pixel
scale is $0.208\arcsec$ per pixel (when binned by 2$\times$2).  

Three 15-sec $V$-band exposures were taken using the FOCAS instrument,
under good seeing conditions of $0.53\arcsec$ FWHM as measured from stars 
in the images. 
The images were bias subtracted and flatfielded using 
standard routines from the IRAF {\tt ccdred} package.  
We then ran the {\tt SExtractor v2.5} code \citep{bertin96}
on the reduced images to generate object catalogs, and we 
matched objects from image to image to determine relative photometric
zeropoints, using {\tt SExtractor} {\tt MAG\_AUTO} magnitudes.
We also astrometrically aligned the world coordinate system (WCS)
of each image to that of the first image, using the IRAF {\tt ccmap} task.
The images were then remapped and coadded, specifically median-combined,
with account made for the relative flux scalings between the images,
using the {\tt swarp v2.16} package\footnote{\url{http://terapix.iap.fr/rubrique.php?id\_rubrique=49}}
The final photometric zeropoint was derived by matching objects detected 
by {\tt SExtractor} in the coadded $V$-band image with those in
the calibrated $g$- and $r$-band images from the APO 3.5m telescope;
the APO images were themselves calibrated using SDSS matches,
as described below in \S\ref{apo_photometry}.
Note this bootstrapping method gives a more robust photometric zeropoint 
as it provides more objects compared to directly matching the 
Subaru and SDSS data, due to the small size of the Subaru image and 
the shallow depth of the SDSS data.
The $g$-band {\tt SExtractor} 3$\arcsec$ aperture magnitudes
were first transformed to $V$-band,
via the relation $V = g - 0.59(g-r) - 0.01$ \citep{jester05}, 
and then used to determine the zeropoint of the Subaru image
using the $V$-band magnitude offsets for matching stars and galaxies
in the images.  The coadded $V$-band image was astrometrically
registered through matches to SDSS objects, again using the 
IRAF {\tt ccmap} task.

\begin{figure}
\plotone{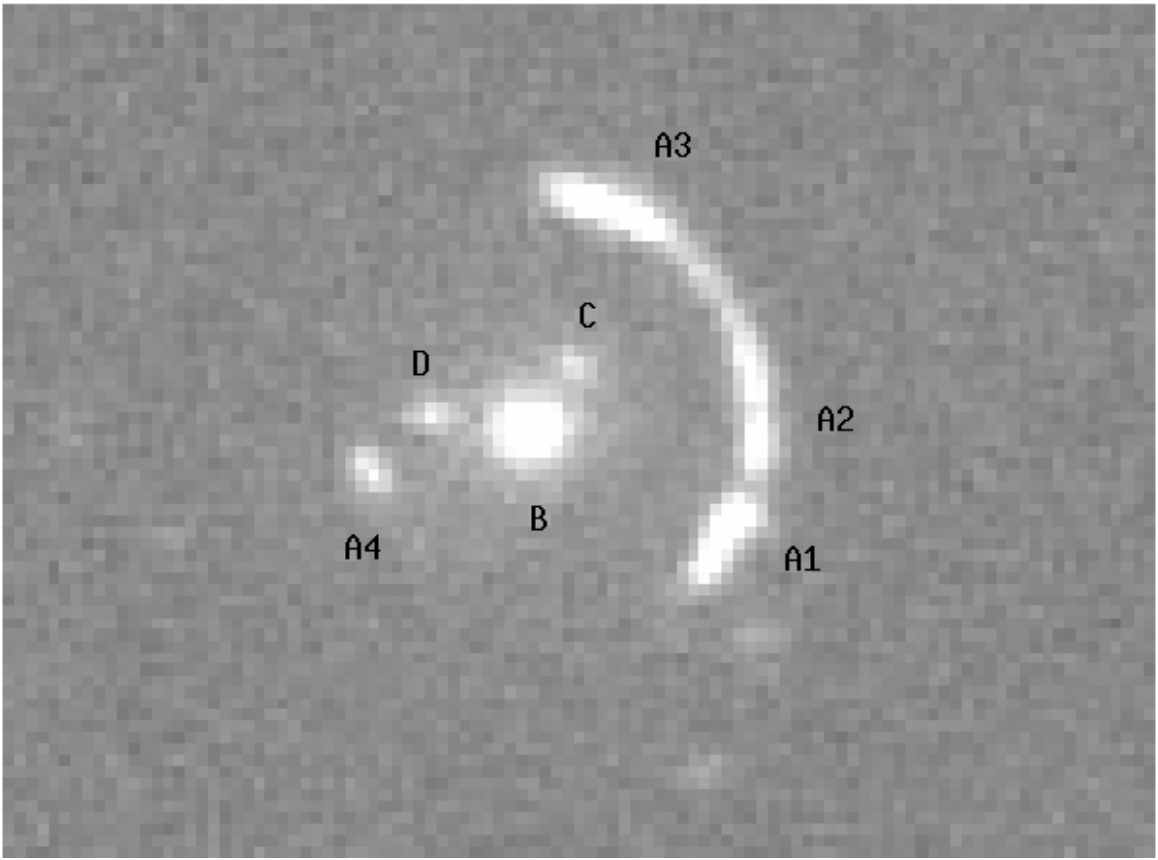}
\caption{The coadded FOCAS $V$-band image. The arc components are
labelled A1, A2, and A3, and the counter-image is A4. The three central
galaxies are labelled B (LRG), C, and D.\label{subaru_image}}
\end{figure}

Fig.~\ref{subaru_image} shows the coadded FOCAS image. Not only is
the counter-image A4 now very clear but we can now see that the central
lensing galaxy (B) is clearly accompanied by two smaller galaxies
(C and D).  Our photometry analysis of this image is described below
in \S\ref{subaru_photometry}.

After the imaging data were obtained, a single 600-sec long-slit 
FOCAS spectrum was also taken, with
the slit oriented to cover both knots A2 and A3 in the arc.
The 300B grating and L600 filter were used, providing a dispersion
of 1.34\AA \ per pixel, spectral coverage of 3700--6000\AA,
and a resolution $R \sim 400$ with a 1.0$\arcsec$-wide slit. 
The Hubble Space Telescope (HST) spectrophotometric standard G191-B2B was
also observed and used for flux calibration.
The FOCAS spectroscopic data were reduced using standard routines from the 
IRAF {\tt twodspec} package.  
The extracted 1D spectra for the A2 and A3 knots are 
shown in Fig.~\ref{arc_spectra}.
The redshift of the arc was found to be $z=2.0010\pm0.0009$
based on measurements of prominent absorption lines due to 
C II, Si IV, Si II, C IV, Fe II, and Al II, typical features seen in 
the spectra of star-forming Lyman break galaxies \citep{shapley03},
in particular of the $z \sim 2$ ``BX/BM'' variety
as defined by the classification scheme of \cite{steidel04}.
Table~\ref{table_spectra} summarizes details about the observed lines.
The high redshift of the knots, combined with the clear arc morphology
seen in the Subaru image, confirm that this is indeed a 
gravitationally lensed system.

\subsection{APO Imaging and Spectroscopy}

Additional follow-up imaging data in the SDSS $griz$ bands were obtained
on the Apache Point Observatory (APO) 3.5m telescope using
the SPIcam CCD imager, which has a scale of 0.28$\arcsec$ per pixel
and a field of view of 4.8$\arcmin \times 4.8\arcmin$.
The data were obtained  under  photometric  conditions, and the
seeing ranged from 0.9$\arcsec$--1.2$\arcsec$.
The total exposure time in each filter was 900~sec, divided into
three dithered exposures (with $15\arcsec$ offsets) of 300~sec each
in order to reject cosmic rays and bad pixels.
Additional details are given in the observation log in 
Table~\ref{table_obslog}.

\begin{figure}
\plotone{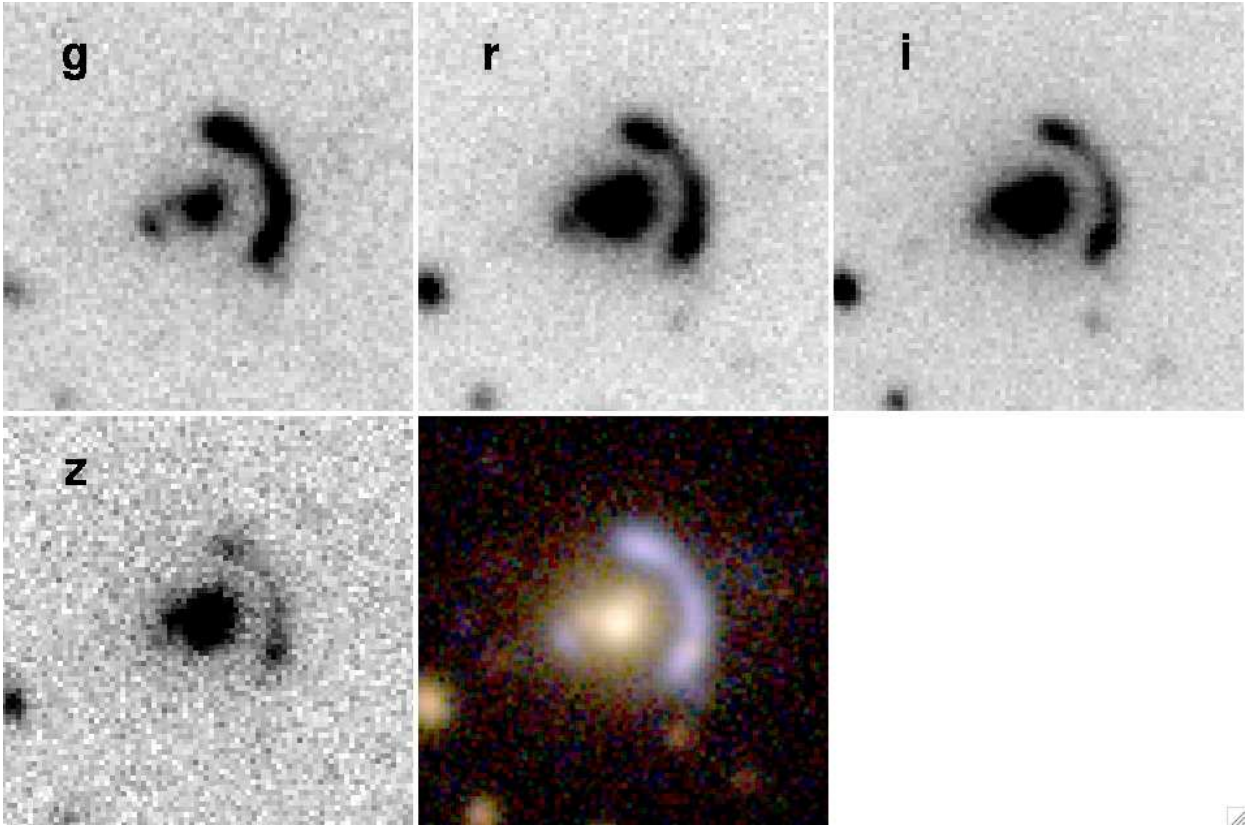}
\caption{The coadded SPIcam $griz$ images and $gri$ color composite.
Note the prominent blue color of the lensed arc and counter-image
contrasted with the redder color of the LRG.
\label{spicam_image}}
\end{figure}

\begin{figure}
\plotone{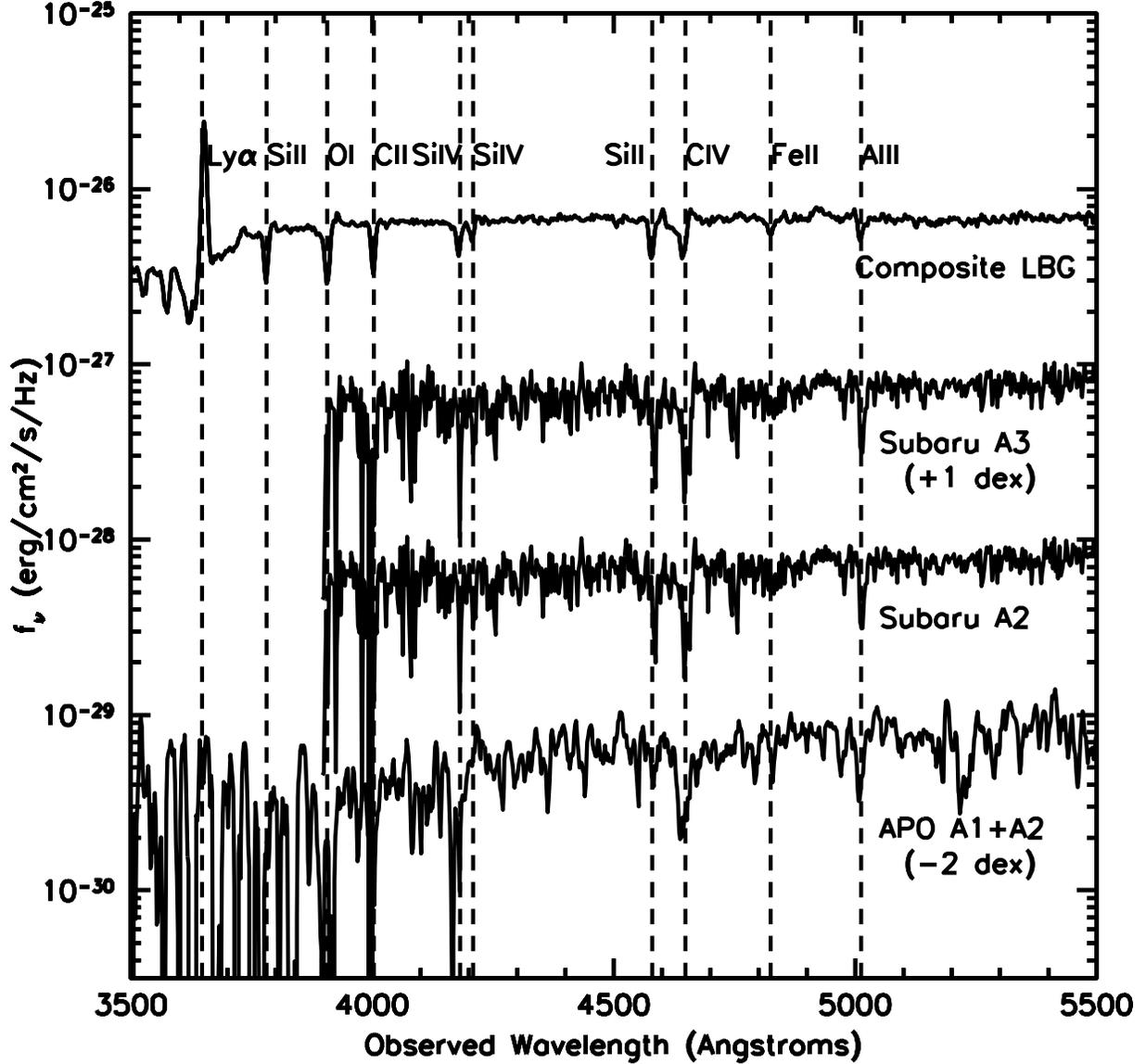}
\caption{Flux-calibrated spectra (in $f_\nu$) for the arc: the Subaru 
spectra show knots A2 and A3 separately, while the APO 3.5m 
spectrum combines knots A1 and A2 and has also been
smoothed (with a boxcar of 5 pixels = 9\AA) to improve S/N.
The spectra have been shifted as indicated in the figure labels
to improve visibility.  A composite LBG spectrum \citep{shapley03}, 
redshifted to $z = 2.00$ to match the arc, is also shown for reference.
The prominent spectroscopic features typical of LBGs are labeled and
indicated by the vertical dashed lines.
\label{arc_spectra}}
\end{figure}

The  resulting $griz$ images were reduced and coadded using the same procedure
described above for the Subaru data.
The SPIcam $z$-band data showed signficant fringing and therefore an 
additional reduction step was necessary to subtract off a master fringe frame.
The final coadded images were again astrometrically registered by matching
to SDSS objects.
The photometric zeropoints for the coadded images were derived using
unsaturated bright stars in the SPIcam images.  Specifically, we 
used GALFIT \citep[][also see below]{peng02} to fit Moffat profiles to
these stars, and compared the resulting total magnitudes to the 
corresponding SDSS model magnitudes.  Note that we did not apply any
color terms in our calibration of SPIcam to SDSS $griz$ magnitudes,
as verified by a comparison of {\tt SExtractor} photometry of the SPIcam
data vs.\ the corresponding SDSS photometry for matching objects.
Fig.~\ref{spicam_image} shows a montage of the coadded
$griz$ SPIcam images, as well as a $gri$ color composite.
We describe our photometry analysis for these images in \S\ref{apo_photometry}
below.  

Additional follow-up long-slit spectroscopy of the arc was carried 
out with the Dual Imaging Spectrograph (DIS III) on the APO 3.5m telescope.
Two 600-sec exposures were obtained, with a 1.5$\arcsec$-wide slit 
covering knots A1 and A2, under $\sim$1.5$\arcsec$ seeing.
The B400/R300 gratings were used, covering an effective
spectral range of 3600--9600\AA, with a dispersion of 
1.83\AA\ per pixel  in  the  blue    part of the    spectrum  and
2.31\AA\  per pixel in the   red.  The spatial  scale is about
0.4$\arcsec$ per pixel.  HeNeAr  lamp    exposures   were taken  for    
wavelength calibration, and the spectrophotometric  standard   
stars GD~50 and  Feige~110 were observed for flux calibration.
The spectra were reduced using the IRAF {\tt ccdred} package 
and the {\tt doslit} task.  The two spectroscopic exposures of the arc
were combined using the {\tt scombine} task, and the red and blue
spectra were spliced together using the {\tt spliceSpec} task from
Gordon Richard's {\tt distools} external IRAF package.  The reduced
spectrum is shown in Fig ~\ref{arc_spectra}.  As with
the Subaru spectra, a redshift was determined from the combined APO
spectrum using absorption features typical of Lyman break galaxies
(see Table~\ref{table_spectra}).  The APO spectrum yields a redshift
of $z=2.0001\pm0.0006$, consistent with that from the Subaru spectra.

\section{Modeling the System}\label{sec:modeling}

\subsection{Subaru Photometry}\label{subaru_photometry}

\begin{figure}
\plottwo{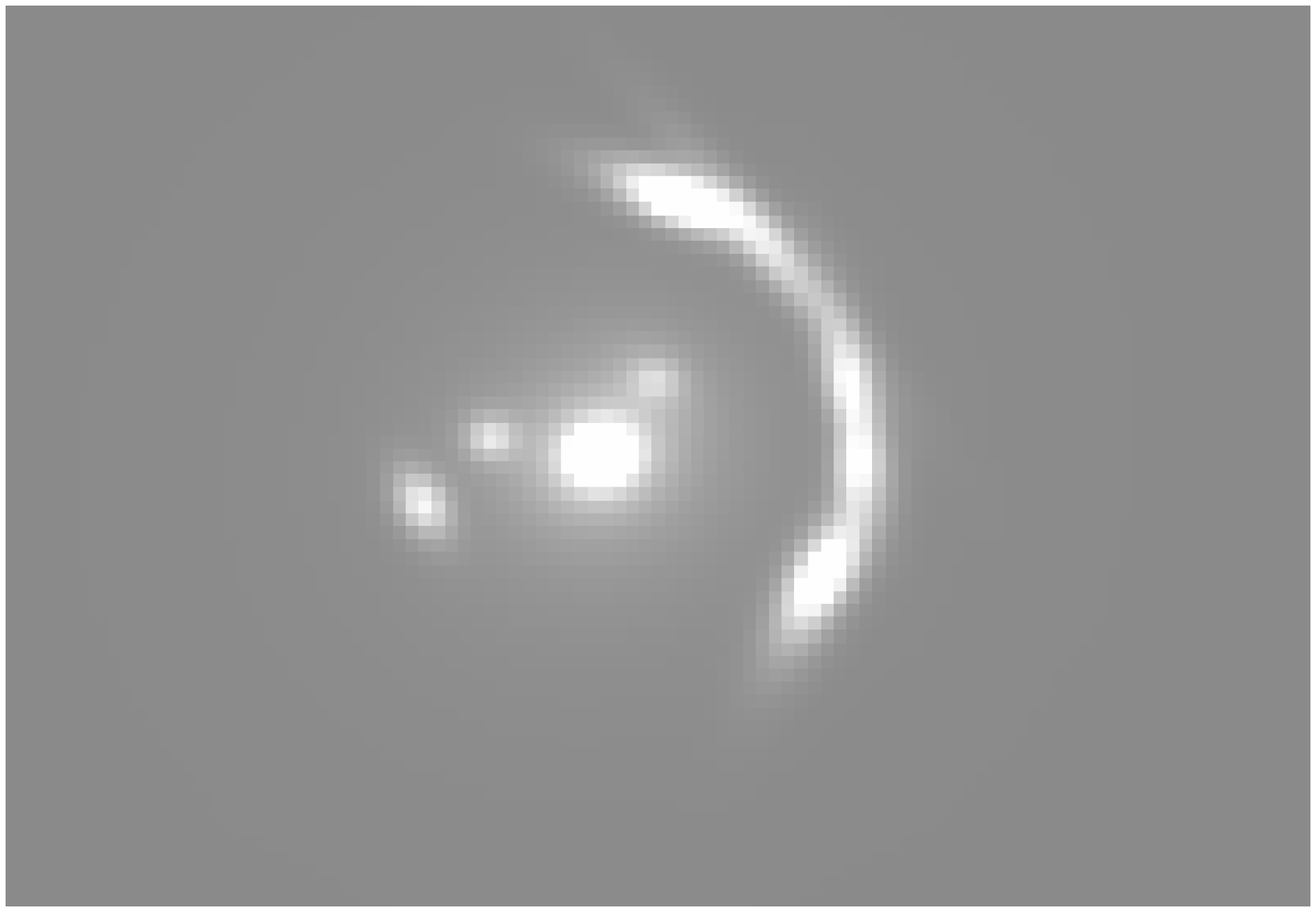}{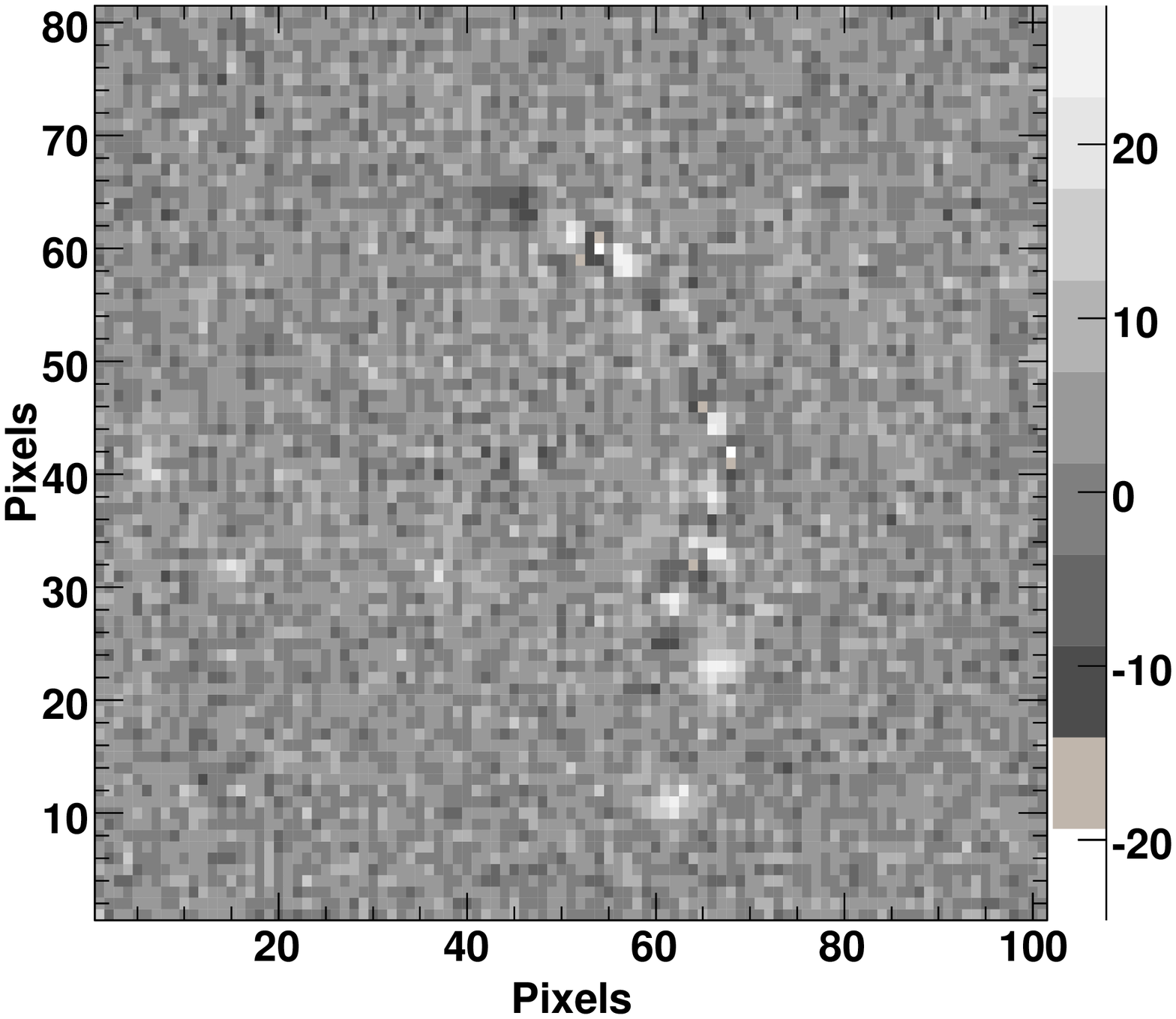}
\caption{The best fit model image ({\it left}) obtained by running 
GALFIT on the Subaru $V$-band image of the system 
and the resulting data-model residual image ({\it right}).
The scale in the right panel is in units of observed counts per 
pixel per 15-second exposure time.
See \S\ref{subaru_photometry} for details.
\label{galfit}}
\end{figure}

We proceed next to derive a lensing model for the Clone system and 
to measure the photometric properties of the lensing galaxies and
the lensed images.
The first step is to model the lens components of the image so that
their light can be subtracted off, leaving us with just the light
of the lensed images that we can use to derive the lensing model,
as described below in \S\ref{lens_modeling}.
To model the lensing galaxies we have used the GALFIT program \citep{peng02}. 
GALFIT can perform a simultaneous fit to multiple objects in a FITS image. 
It allows the user to fit a number of common galaxy profiles such as
Sersic, de Vaucouleurs, and exponential disk. The inputs required are a
FITS image of the system, a FITS file of the point spread function (PSF), 
an optional mask
which can be used to eliminate pixels from consideration in the fit,
and a determination of the sky background. The initial object
positions were determined using SExtractor. 
The modeling was done using the coadded $V$-band Subaru image
as it has the highest resolution. The PSF was determined from stars in
the image. We also included the arc and counter-image in the
GALFIT model, but did not include the two faint galaxies that
can be seen in the bottom right of Fig ~\ref{subaru_image}.
The best description of the system is obtained using a Sersic profile 
for the main LRG, de Vaucouleurs profiles for the two small galaxies
(C and D),
and a combination of 5 exponential disks for the arc and one
exponential disk for the 
counter-image. This gives a $\chi^{2}/{\rm dof}$ of 1.13.
In Table~\ref{galfit_table} we show the fitted parameters and in
Fig.~\ref{galfit} we show the model and the data--model residual image. 
From the residual image we can see that the 
galaxies B, C, and D are well modeled, but that the exponential disk
model for the arcs is not perfect.  We then subtract off the models for 
just the lens objects B, C, and D from the image, leaving us with the light 
of the lensed arc and counter-image for the subsequent lens modeling.

\subsection{Lens Modeling} \label{lens_modeling}

\begin{figure}
\plottwo{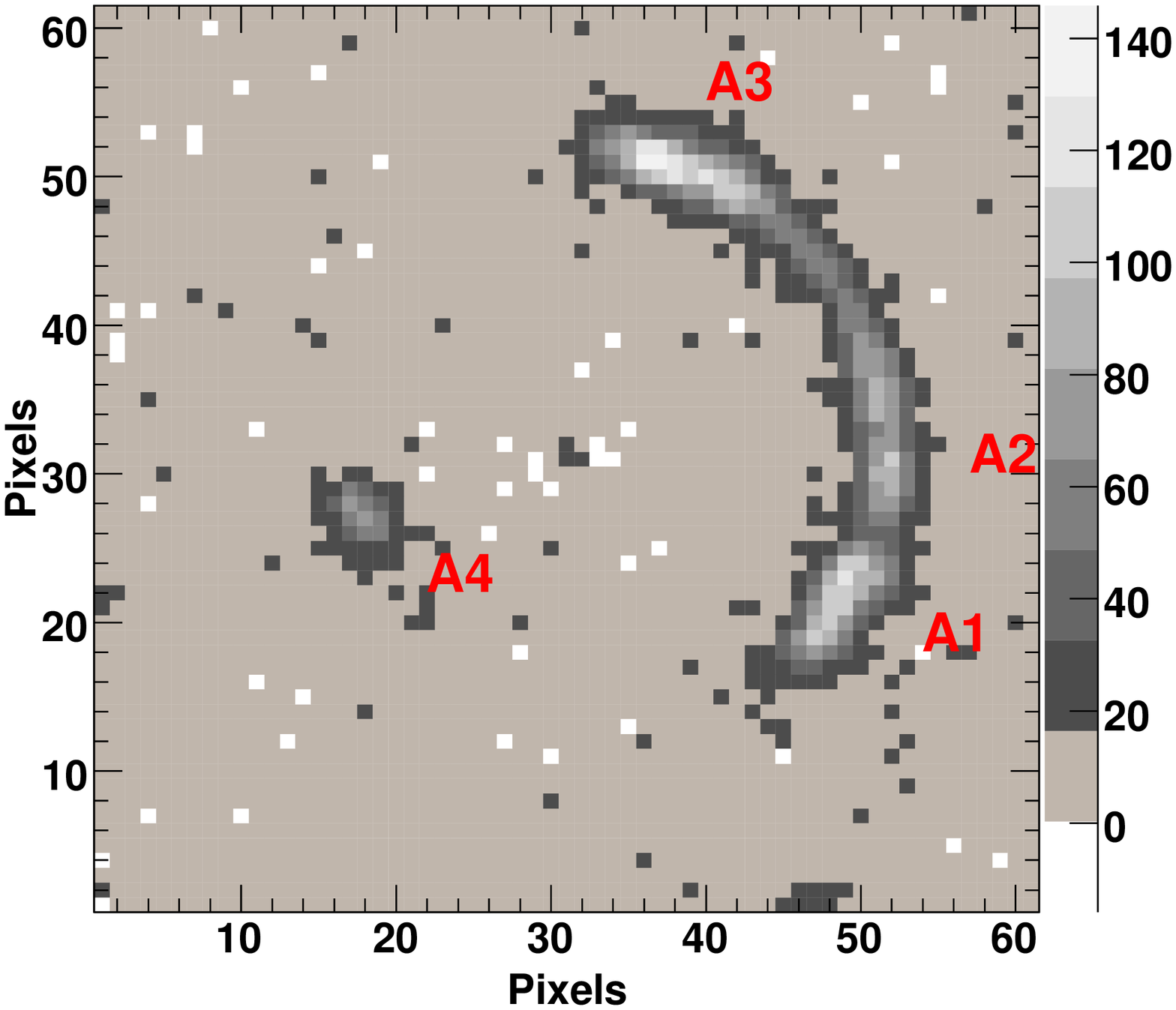}{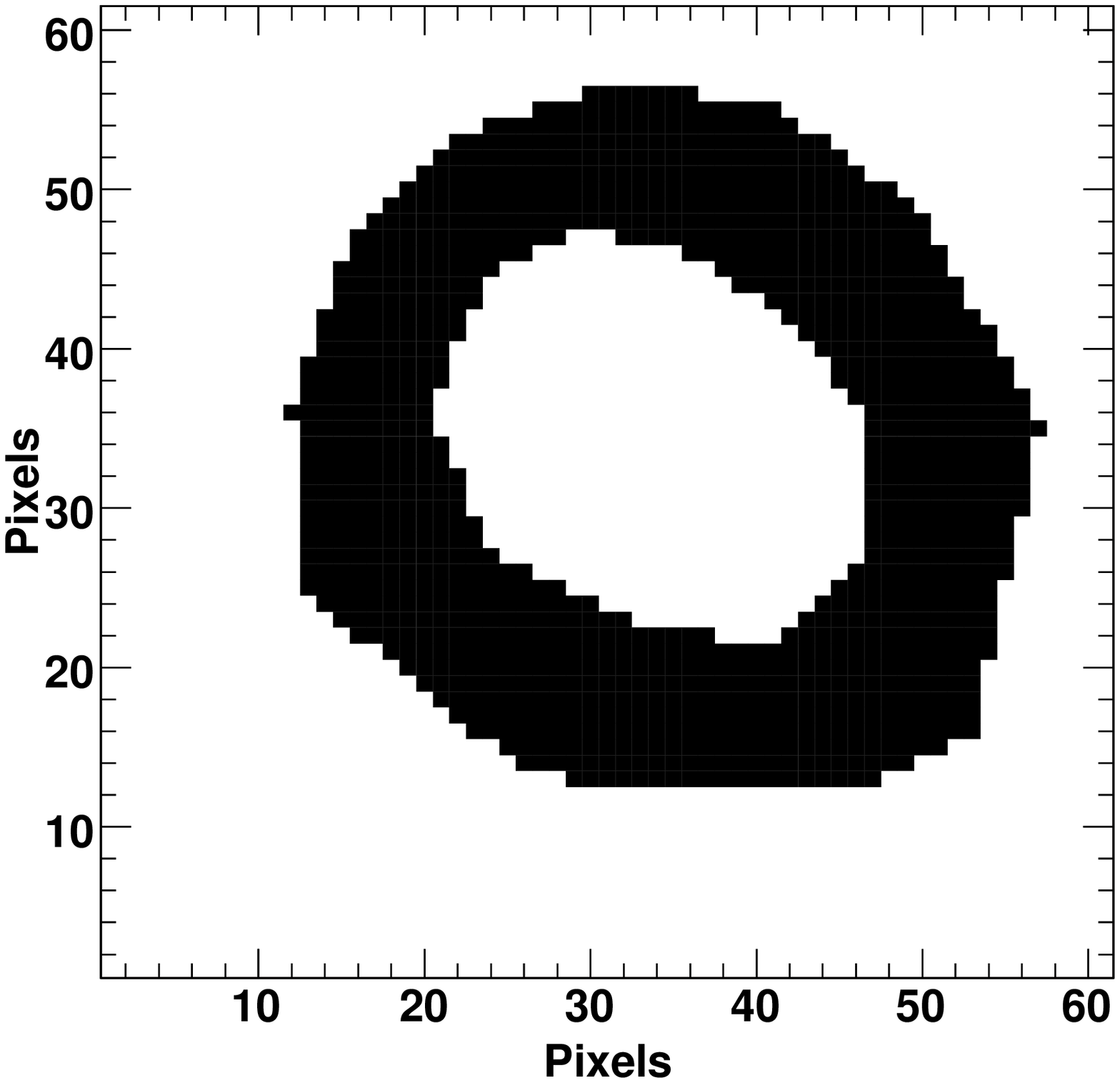}
\caption{The input image ({\it left}) and the pixel mask ({\it right}) 
used in the LENSVIEW lens model fits. 
LENSVIEW will use only the pixels inside the mask (the black region
in the right panel) to calculate the $\chi^{2}$ between the input image 
and the model.  The scale in the left panel is in units of observed 
counts per pixel per 15-second exposure for the Subaru $V$-band image.
See \S\ref{lens_modeling} for details.
\label{lens_input_image_mask}}
\end{figure}

We have modeled the lens using the LENSVIEW program
\citep{wayth06}, a program for modeling resolved
gravitational lenses. It is based on the LENSMEM algorithm
\citep{wallington96} and uses a
maximum entropy constraint to find the best fitting lens mass model and
source brightness distribution. It supports a number of common mass models.
The inputs to the program are a FITS image of the lensing system with
the non-arc objects removed (Fig.~\ref{lens_input_image_mask}, left plot), 
a FITS file containing the PSF for the image, 
a FITS image of the pixel-by-pixel variance of the data, an empty FITS
image with the dimensions of the desired source plane,
and a FITS image containing a mask of the pixels over which the
$\chi^{2}$ will be calculated (Fig.~\ref{lens_input_image_mask}, right
plot). 
It also requires the ratio of the angular size of the
pixels between the image and source planes.
We have used a source plane of $10 \times 10$ pixels,
with $0.052\arcsec$ per pixel, i.e., 4 times finer than the image
plane pixel scale.

\begin{figure}
\plottwo{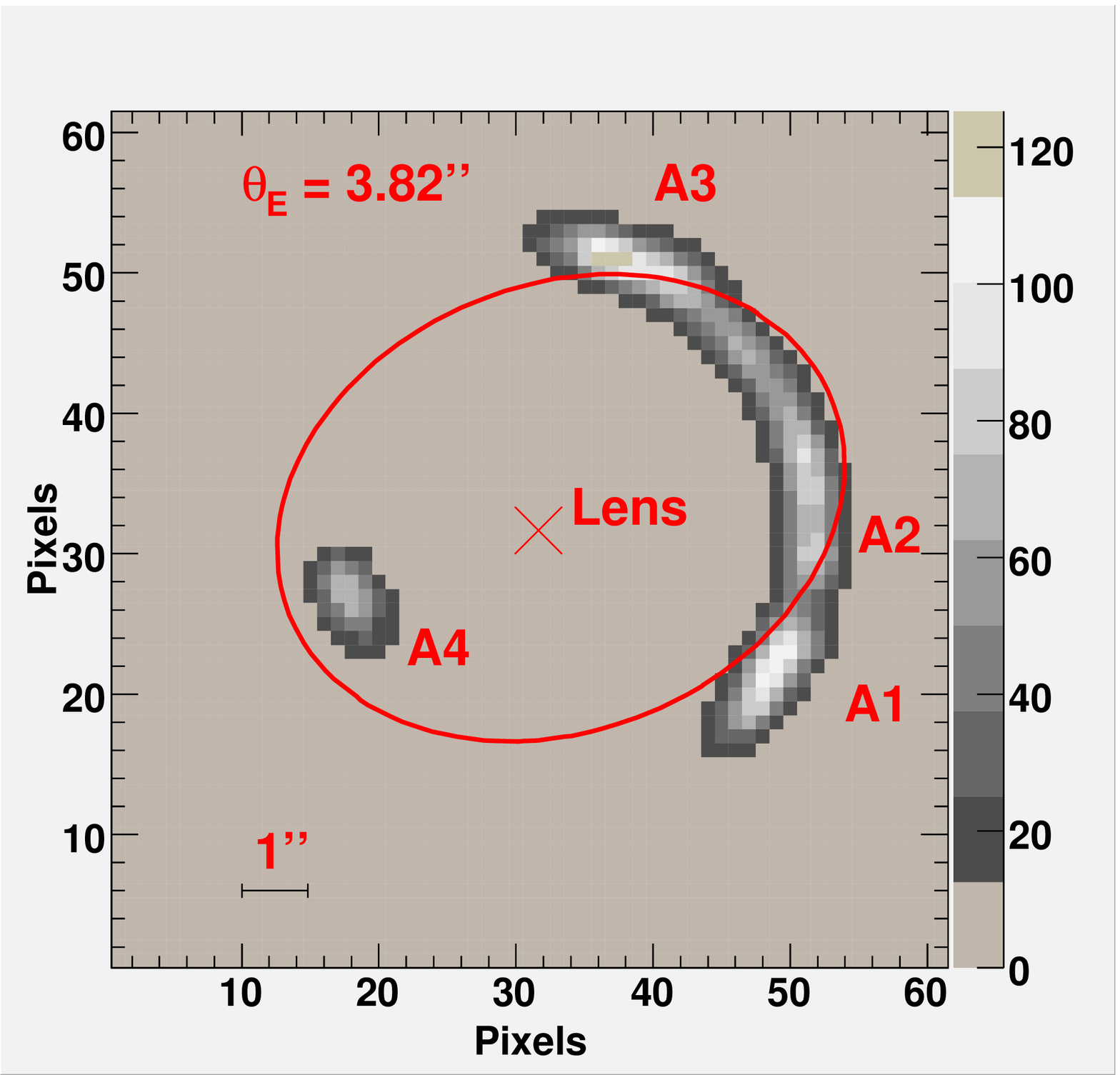}{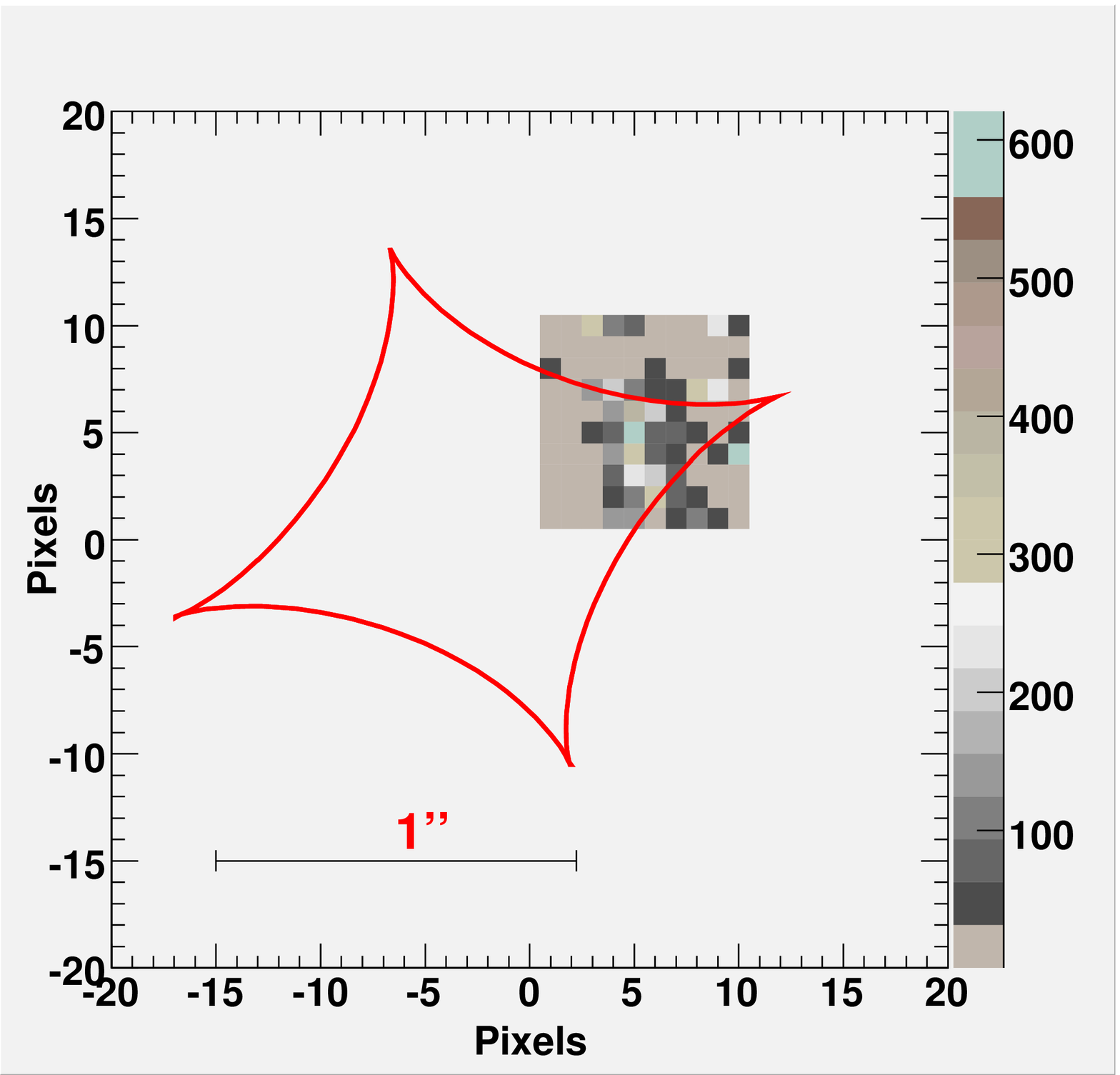}
\caption{The best fit LENSVIEW model image ({\it left}). The
tangential critical line is shown in red. The best
fit LENSVIEW model source ({\it right}). The tangential caustic is
shown in red along with the $10\times 10$ pixel source plane. On both
plots the spatial scale is indicated by the horizontal bar representing
$1\arcsec$.  The flux scales in the panels are in units of observed 
counts per {\it image plane} pixel ($0.208\arcsec$~pixel$^{-1}$)
per 15-second exposure for the Subaru $V$-band image.
Note the source plane pixels ($0.052\arcsec$~pixel$^{-1}$)
in the right panel are 16 times smaller in area than the 
image plane pixels in the left panel.
\label{lens_model_images}}
\end{figure}

\begin{figure}
\plotone{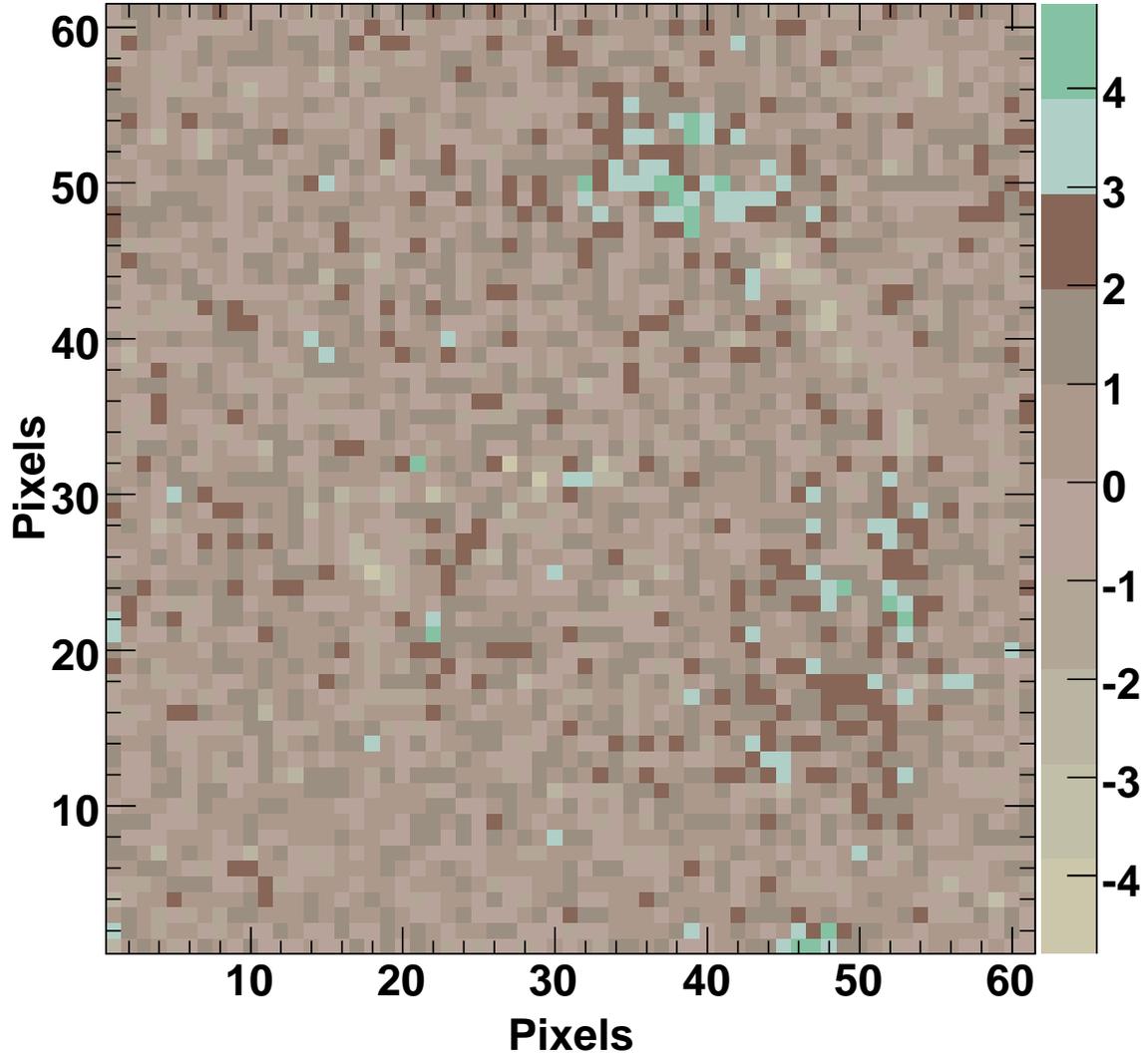}
\caption{The normalized residual image, 
$({\rm counts}_{data} - {\rm counts}_{model})/\sigma_{data}$,
for the best fit LENSVIEW model of the system.
\label{lens_model_residual}}
\end{figure}

Using LENSVIEW we have modeled the system using a singular isothermal
ellipsoid \citep[SIE;][]{kormann94} as the mass model. 
The best fit model yields an Einstein  radius of 
$\theta_{\rm Ein} = 3.82\pm{0.03}\arcsec$,
which translates to $R_{\rm  Ein} = 14.8\pm{0.1}\ h^{-1}$~kpc at the 
LRG redshift of 0.422.
The  fitted axis ratio and position  angle  are $0.751\pm{0.018}$ and 
$-70.11\pm{0.39}\arcdeg$ (E of N), respectively.
The best fit model with
the tangential critical curve is shown in Fig.~\ref{lens_model_images},
left plot, and
the predicted source with the corresponding tangential caustic is given in
Fig.~\ref{lens_model_images}, right plot. The best-fit lens center is offset
by a small amount, ($0.07\arcsec,0.04\arcsec)$ in (RA,Dec), which is
much less than 1 pixel (recall the scale is $0.208\arcsec$ per pixel)
from the center of the LRG light distribution obtained from GALFIT.
The total magnification of the system, obtained by dividing the total 
flux in the arcs by the total flux in the source, is $27 \pm 1$. 
Comparing Fig.~\ref{lens_input_image_mask} (left)
and Fig.~\ref{lens_model_images} (left), we see that the model does
look qualitatively quite like the data. 
The best fit  $\chi^{2}/{\rm dof}$ is 2.18 (2102 for 968 dof), however,
indicating formally a poor fit.  This can be understood by looking at
the pixel-by-pixel residuals scaled by the errors,
$({\rm counts}_{data} - {\rm counts}_{model})/\sigma_{data}$, shown in
Fig.~\ref{lens_model_residual}.
We see that there are large residuals coming from the A3 knot, which
is brighter in the data than in the model by 23\% within a $3\arcsec$ aperture. 
We have explored other mass models including SIE+external shear but 
find similar or worse agreement. 

In strong lensing it has been known for 
some years that the
smooth mass models fit the image positions well but not always the flux
ratios of the images.  
As LENSVIEW uses the full image information it is not
possible to use it to determine how well the image positions alone are
determined. So we turn to gravlens/lensmodel \citep{keeton01} which allows us to fit
an SIE model using only the image positions. 
We use the A1-A4 image positions determined by running {\tt SExtractor}
on the Subaru image and given in Table~\ref{table_photometry} 
(same as used below in \S\ref{apo_photometry}).
We assign large errors to the flux
ratios so that they do not contribute to the $\chi^{2}$. We obtain a very good fit to the image positions, with a
$\chi^{2}$ of 2.55 for 3 dof and values of the SIE parameters that agree
with those from the LENSVIEW fit. As the image positions are well
determined, the statistical errors quoted above for our lens model
parameters are from the lensmodel fit rather than the LENSVIEW fit.
The predicted flux for A3 in the lensmodel fit is smaller than
the measured flux by a factor of 2. This is more discrepant than what we 
obtained from LENSVIEW above, in which the source light distribution 
is more realistically modeled as an extended source, as opposed to a 
point source as used in lensmodel.  The A3 flux is also not better matched
by adding external shear or by adding galaxies C and D as singular 
isothermal spheres.  An interesting discussion of anomalous flux ratios
in 4-image lenses with a fold configuration, as is the case for our system,
can be found in \cite{keeton05}.  They define the ratio $R_{fold} =
(F_{+} - F_{-})/(F_{+} + F_{-})$, where $F_{+}$ and $F_{-}$ 
are the observed fluxes for a pair of images of opposite parity, as
indicated by the subscripts.
They model $R_{fold}$ for different image pairs in
4-image lenses. Deviations from the expected values are thought to
indicate the presence of structure at scales smaller than the
separation between the images. We measure $R_{fold} = 0.173$ for the
image pair A3-A2. This value is not consistent with the range of
values shown in Fig.~5 of \cite{keeton05}. 
Given this result and our poor $\chi^{2}$ from LENSVIEW we conclude that 
we may have substructure in the lens which is currently not being 
well modeled using a smooth SIE mass distribution.


From the SIE model, the   velocity  dispersion  of  the   mass
distribution  doing   the lensing  is
$440\pm{7}\ {\rm km s}^{-1}$, which would be quite large for an 
elliptical galaxy.  The SDSS database does not provide a spectroscopic
velocity dispersion for the LRG due to the low signal-to-noise of the 
SDSS spectrum.  We obtained a similarly large value for the velocity 
dispersion of the 8~o'clock arc lensing mass, which is discussed in 
\citet{allam07}.  Combined with the large $3.82\arcsec$ Einstein radius
and the presence of neighboring red galaxies like C, D, E 
(see \S\ref{apo_photometry}),
and others further away, this indicates that the lensing is due in part to the 
group environment around the central LRG \citep[see, e.g.,][]{oguri06}.
We have thus investigated two alternative mass models to attempt a
better approximation of the group lensing contribution, specifically
using SIE+external shear and a Navarro, Frenk, \& White 
(NFW) profile \citep{navarro97}.  However, the LENSVIEW fits in both
cases give about 10\% worse $\chi^2$ per dof than the simple SIE model,
and in particular the SIE+external shear model gives only a small 
shear of 0.006 that is closely aligned with the position angle of the 
main SIE profile.  We will further investigate
the group lensing environment, as well as the substructure issues noted
above, using higher-resolution HST imaging data that we are analyzing
for this system \citep{hstpaper}.  Nonetheless, using the current data 
and a simple SIE fit, we are able to provide a reasonable
model that reproduces the most salient features of the lensing system, 
namely the positions and morphology of the lensed arc and counter-image.

Since both the redshift of the LRG and the source are known we are able
to determine the angular diameter distance to the source ($D_{s}$), to
the lens ($D_{l}$), and between the source and  lens ($D_{sl}$), to be
1209, 801 and 829 $h^{-1}$ Mpc, respectively.  Then
from  the simple SIE  model we  can  determine the  mass interior to
$R_{\rm  Ein}$  using   $M_{\rm Ein}=  (c^{2}/4G)(D_{l}  D_{s}/D_{sl})
\times   \theta_{\rm Ein}^{2} = 2.10\pm{0.03}\times10^{12} \ h^{-1} M_{\sun}$.  
As we are using the SIE convention of \citet{kormann94}, to be more precise
the enclosed mass is actually defined within an elliptical aperture 
with semi-major axis $\theta_{\rm Ein}$, and axis ratio and position angle 
as given above.  For the same aperture, we also determine the lens light,
by summing the fluxes from the best-fitting GALFIT models for the LRG and for
galaxies C and D (see \S~\ref{subaru_photometry} and \ref{apo_photometry});
the results are given in Table~\ref{table_photometry}.  Note that due to 
the similarity of the Einstein radius of the lens mass model to the 
half-light (or effective) radius of the LRG, and likewise for the 
respective axis ratios and position angles (cf.\ Table~\ref{galfit_table}), 
the flux within the lens light aperture is very close to half the total
flux of the LRG (galaxies C and D contribute only a small amount).  
We then convert the apparent lens light to absolute fluxes,
adopting $k$-corrections using an elliptical galaxy template \citep{coleman80},
and obtain mass-to-light ratios in the rest-frame $gVriz$ bands of
$M/L = 27, 22, 19, 15$, and 12~$h \ M_{\sun}/L_{\sun}$, respectively
($\Omega_M = 0.3, \Omega_\Lambda = 0.7$).
We note that these $M/L$ values, out to a radius of $15~h^{-1}$~kpc,
are $\sim$5-10 times larger than those for the lensing LRGs,
on the scale of a few kpc, from the SLACS sample \citep{treu06,koopmans06}.
As shown Fig.~7.8 of \cite{kubik07}, this trend of $M/L$ with radius
is consistent with that determined for elliptical galaxies
using independent dynamical and X-ray techniques \citep{bahcall95}.

\subsection{APO Photometry}\label{apo_photometry}

\begin{figure}
\plottwo{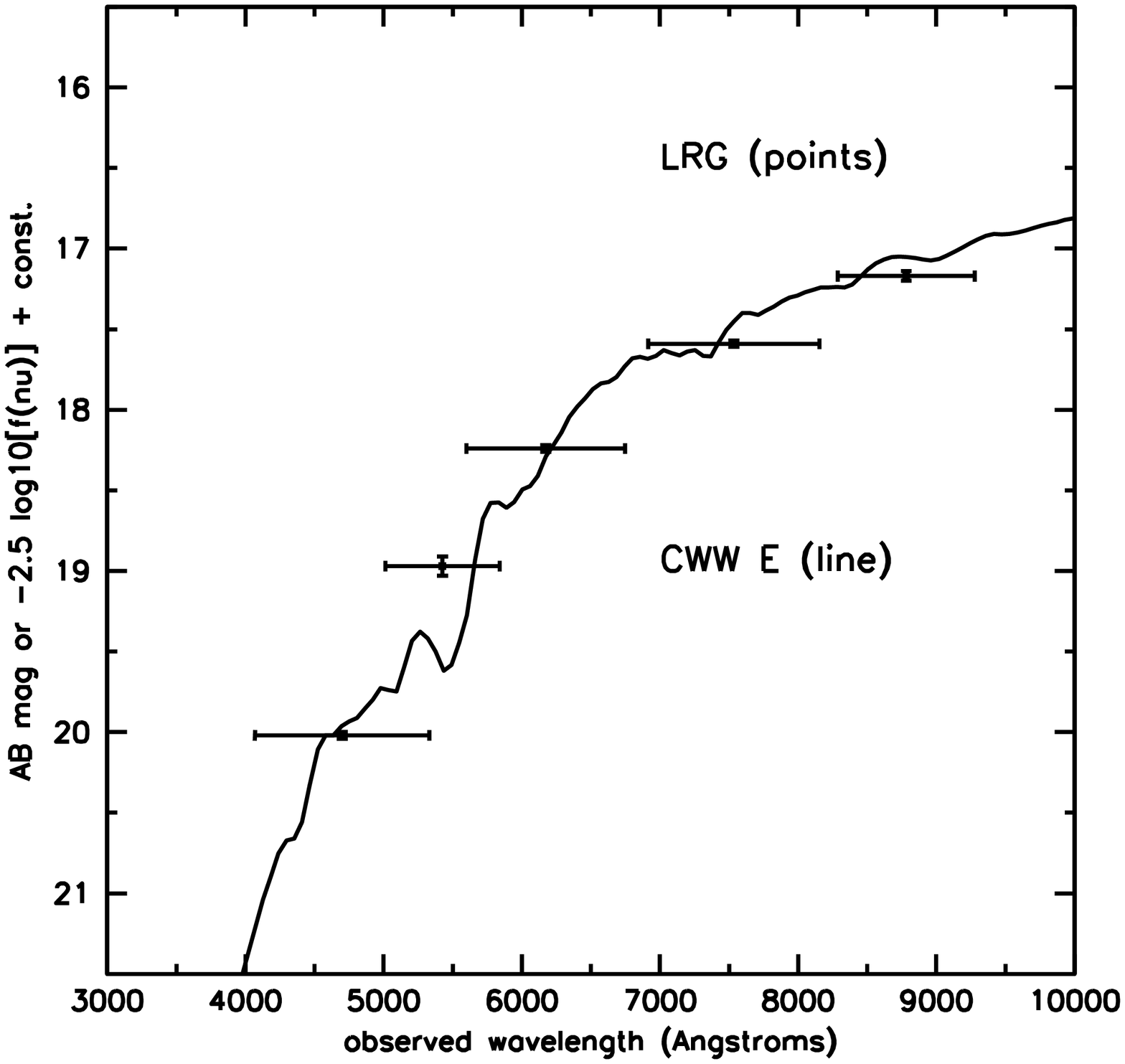}{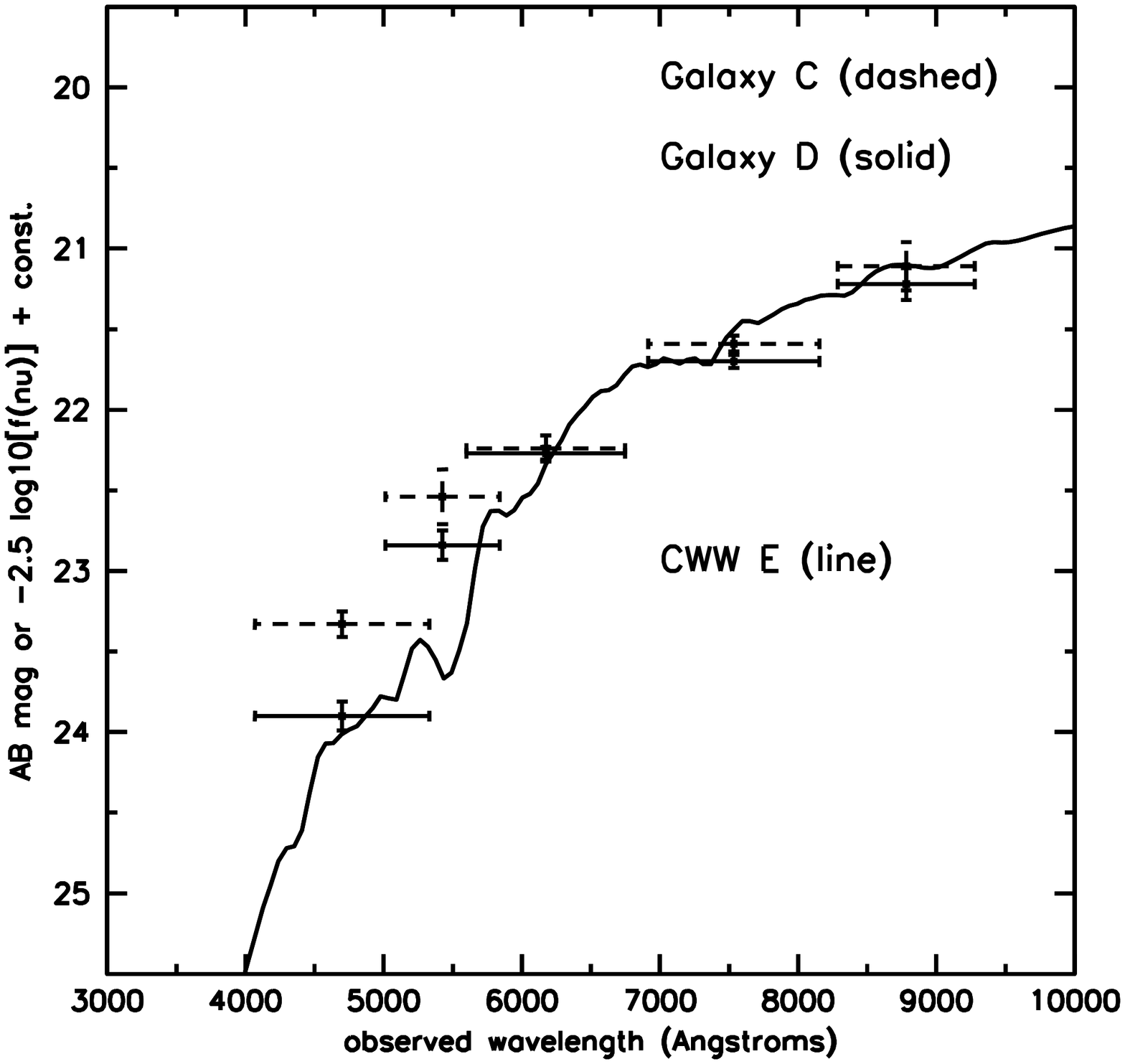}
\caption{The $gVriz$ AB magnitudes ({\it points with error bars})
of the LRG ({\it left}) and 
galaxies C and D ({\it right}) are compared to the 
rescaled spectrum of a template elliptical galaxy (CWW E) 
from \cite{coleman80}.  The CWW E template spectrum has been redshifted
to the LRG redshift $z = 0.422$.
\label{lensing_galaxy_magnitudes}}
\end{figure}

We turn now to the photometry analysis of the APO 3.5m SPIcam coadded imaging
data in order to derive color information for the various lensing galaxies and 
lensed image components.  Because the SPIcam data were taken
under only modest seeing conditions, we will rely on the galaxy 
profile parameters determined earlier from running GALFIT on 
the Subaru $V$-band image, rather than try to re-fit those parameters
independently in each of the SPIcam $griz$ images.
Specifically, we adopt all the best-fit $V$-band profile parameters for 
the LRG (= galaxy B), galaxies C and D, and counter-image A4, {\it except}
that we will fit for the total magnitude of each of those four components.
Moreover, we also re-fit for the position of the LRG, in order to
account for small errors in the astrometric registration relative to
the Subaru image;  we find best-fit shifts of
$\leq 0.07\arcsec$, which are small but nonetheless result
in noticeable visual improvement in the residual image after subtracting 
off the LRG model.  Note we do not attempt to fit models to the lensed arc 
images, as was done for the Subaru data.  Instead, we mask out
the image areas corresponding to the A1, A2, and A3 components before
running GALFIT on the SPIcam data.  The masks are derived using
{\tt SExtractor}-generated ``segmentation'' images, which flag the pixels
belonging to each detected object.  We will later compute aperture magnitudes
for the arc components, in a model-independent way as described below.  
For the PSF model needed by GALFIT, we use the best-fit Moffat profile 
derived by GALFIT for a bright unsaturated star in a given image.  
We find that our results are not sensitive to whether we use the 
Moffat profile or the actual data for the star itself as the PSF model.
Note we also first use {\tt SExtractor} to do sky subtraction on an
image before feeding it to GALFIT.  Our GALFIT photometry results for the 
SPIcam coadded $griz$ images are given in Table~\ref{table_photometry}.
We plot the $gVriz$ total magnitudes of the LRG and of
galaxies C and D in Fig.~\ref{lensing_galaxy_magnitudes},
where we have also overlaid a template elliptical galaxy spectrum from
\citet{coleman80}, after redshifting to the LRG redshift $z = 0.422$
and converting the flux of the spectrum to AB magnitude units.
The reasonable match of the spectral energy distributions (SEDs; described by 
the $gVriz$ magnitudes) of galaxies C and D to the template spectrum 
is consistent with the interpretation of those two galaxies as
early-type galaxies at the same redshift as the LRG.

As noted above, for the lensed arc image components A1-A3, 
we measure simple aperture magnitudes.  We do this instead of attempting 
profile fitting since we do not expect the lensed and distorted 
arc images to follow standard galaxy profiles, as can be seen in the
residual image shown in Figure~\ref{galfit} (right panel) for the Subaru data.
We measure 3$\arcsec$-diameter circular aperture magnitudes for
each of the A1-A3 arc components, with centers determined
from running {\tt SExtractor} on the Subaru image. 
The aperture magnitudes are measured from the images after subtraction of
the best-fit GALFIT galaxy models as described above.
The Subaru $V$-band image is first convolved by a Gaussian to degrade the 
seeing to 1.0$\arcsec$ to match the typical seeing in the 
SPIcam data.  Otherwise no aperture
corrections are made to reconcile the small seeing differences among
the $griz$ data.  We also ignore a small overlap in the apertures
centered on the A1 and A2 components and do not attempt any deblending.
In addition, we define a partial annular aperture, centered on 
the LRG, with inner radius 3$\arcsec$, outer radius 5.5$\arcsec$, 
and position angle ranging from $-140$ to $+5$ degrees E of N.
This (partial) annulus provides a simple aperture that captures the shape and 
flux of the lensed arc.  Our aperture photometry results for lensed arc
images are presented in Table~\ref{table_photometry}.

\begin{figure}
\plotone{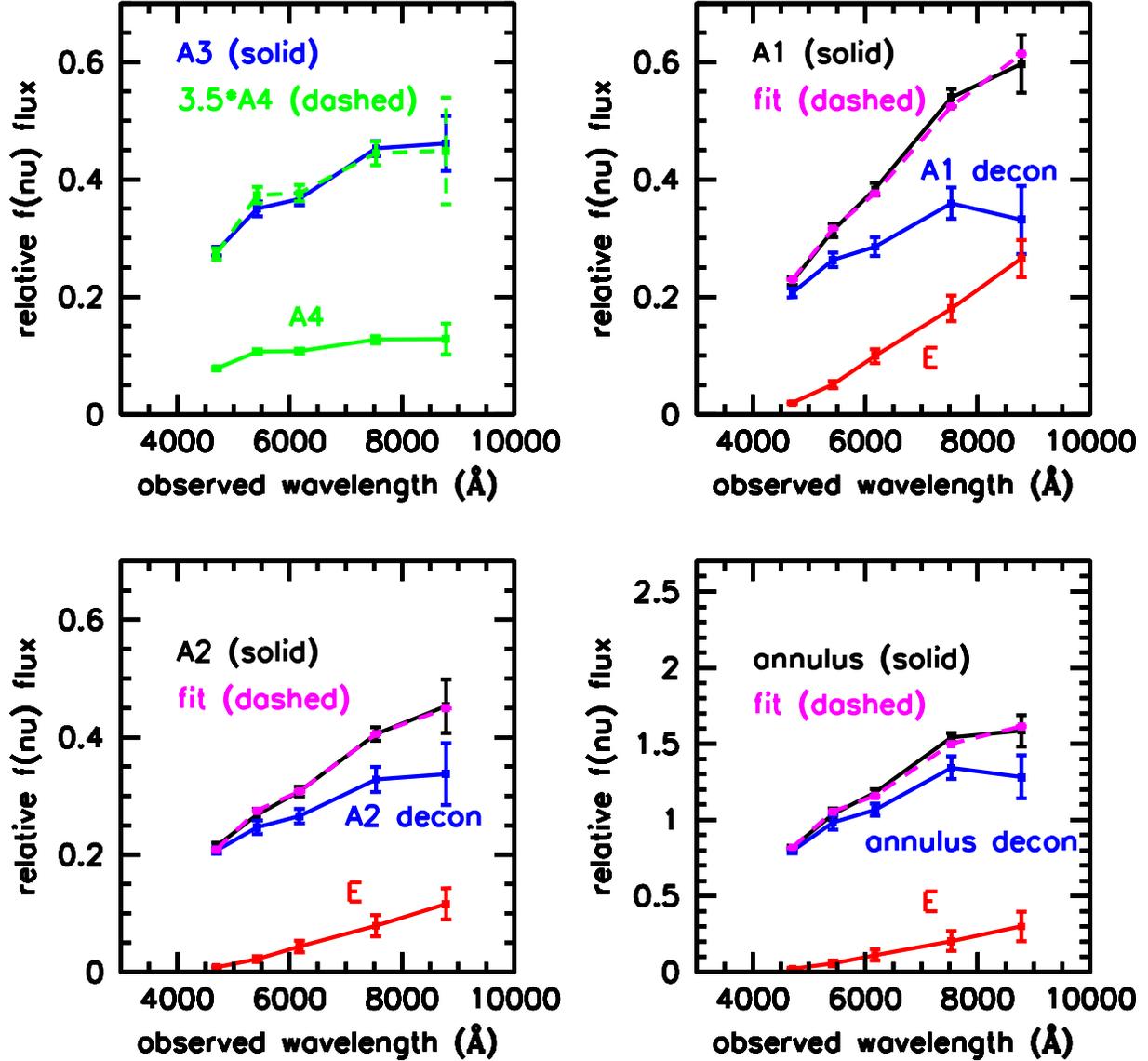}
\caption{The $gVriz$ magnitudes for the lensed image components A1-A4,
for a partial annular aperture (``annulus'') containing the lensed arc, 
and for galaxy E are shown.  Results are also shown after a fitting 
procedure is used to decontaminate (``decon'') the light of galaxy E from 
the A1, A2, and partial annulus apertures.  Please see \S\ref{apo_photometry}
for the full details.  Note the vertical scale is in arbitrary, relative
units that are {\it linear} in the $f_\nu$ flux.
\label{lensed_image_magnitudes}}
\end{figure}

We see from Fig.~\ref{lensed_image_magnitudes} that the 
lensed A3 component and the counter-image A4 show excellent agreement
in their SEDs, as described by their $gVriz$ magnitudes.
This also gives us confidence that our GALFIT galaxy fitting and
model subtraction procedure is working well, as A4 is 
significantly fainter than A3 and one might have expected that 
A4's photometry is prone to proportionately more error due
to contamination by the light of the LRG.
However, Fig.~\ref{lensed_image_magnitudes} also shows that the A2 
and especially the A1 components are significantly redder than the A3
component.  This may also be seen from Fig.~\ref{spicam_image}, where
the upper part of the A1 knot appears noticeably redder than the rest 
of the arc.  It turns out from our higher-resolution HST data \citep{hstpaper}
that there is also a small red galaxy, henceforth galaxy E, 
inside A1's 3$\arcsec$ aperture.  There is also some flux
from galaxy E contaminating the A2 aperture due to seeing.
Although galaxy E is not resolved from the 
arc even in our best-seeing ground-based data, the Subaru image,
we can nonetheless infer its brightness by assuming it has the 
same SED as the LRG.  We therefore decompose the flux within the A1, A2, 
and annular apertures as a linear combination of two flux components:
one with the SED of A3 to account for the lensed source galaxy flux,
and the other with the SED of the LRG to account for the galaxy E flux.
Doing this allows us to ``decontaminate'' the galaxy E flux from
the A1, A2, and annular apertures.  
As shown in Fig.~\ref{lensed_image_magnitudes},
this procedure produces good fits to the total fluxes within the
A1, A2, and annular apertures, and in that figure and in 
Table~\ref{table_photometry} we also show the ``decontaminated'' 
fluxes obtained by subtracting the best-fit LRG-component flux from the
original fluxes within each of those apertures.
Note that this is not a trivial result as the procedure uses only
two parameters (the two flux components) to fit the five data points
(the $gVriz$ fluxes) for each aperture.  
Also, from the annulus result we find that the
galaxy E flux is about 2\% of the LRG flux, making galaxy E comparable
to galaxies C and D in brightness (see Table~\ref{table_photometry}).
In the $V$-band, the galaxy E flux contaminating the annular aperture is 
about $6\%$ of the lensed source galaxy flux,
implying only a small perturbation for the LENSVIEW 
modeling results presented above using the Subaru image.

\section{Source Galaxy Star Formation Rate}\label{sec:SFR}

\begin{figure}
\plotone{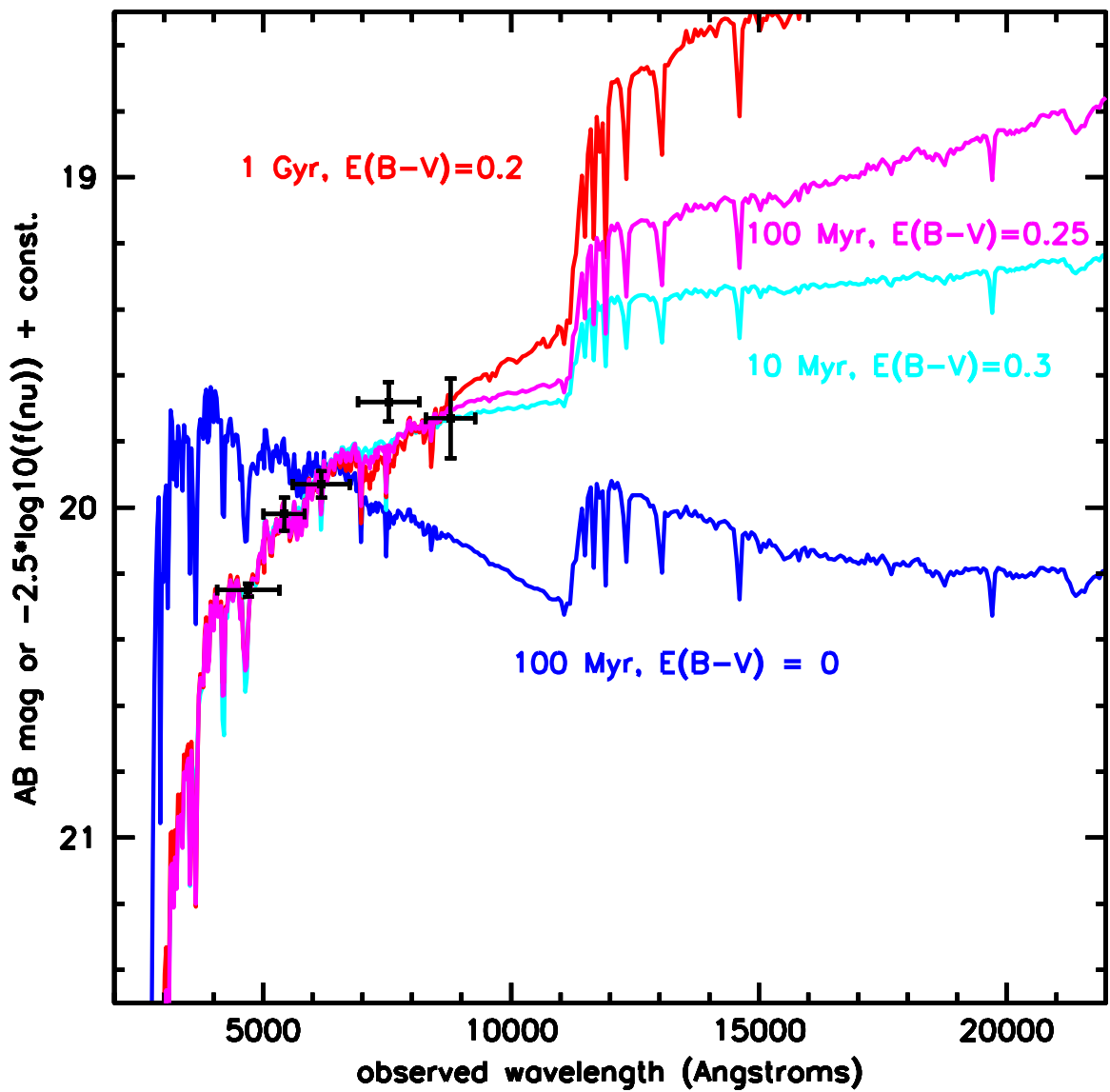}
\caption{The $gVriz$ magnitudes ({\it black points with error bars})
of the arc in the partial annulus aperture are compared to 
four different \cite{bc03} constant star formation rate models normalized
to the data.  The models span a range of different ages and dust extinction
values:
100 Myr, $E(B-V) = 0$ ({\it blue}); 
 10 Myr, $E(B-V) = 0.3$ ({\it cyan}); 
100 Myr, $E(B-V) = 0.25$ ({\it magenta}); and
  1 Gyr, $E(B-V) = 0.2$ ({\it red}).
See \S\ref{sec:SFR} for details.
\label{SED_models}}
\end{figure}

To estimate the star formation rate (SFR) and the amount of dust extinction, 
we compare our $gVriz$ SED data to models from the 
GALAXEV \citep{bc03} stellar population evolution package.
We use simple constant star formation rate models,
with solar metallicity and a \citet{salpeter55} initial mass function (IMF) 
over the mass range $0.1-100 \ M_\sun$, and we also add dust extinction
according to the prescription of \citet{calzetti00}.  
We consider ages of 10 Myr, 100 Myr, and 1 Gyr for the models,
and set values of the (stellar continuum) color excess $E(B-V)$ 
to 0.3, 0.25, and 0.2, respectively, in order to get a good visual match 
to the slope of the $gVriz$ data for the annulus aperture; 
see Fig.~\ref{SED_models}.
We set the normalization of each model by finding the average
offset between the observed and model $gVriz$ magnitudes.
To facilitate comparison with previous simple SFR estimates for
cB58 and the 8 o'clock arc, we express our results in analogy 
with eqn.~(6) of \citet{pettini00}, but using a flat cosmology with 
$\Omega_M = 0.3, \Omega_\Lambda = 0.7$, 
and $H_0 = 100 h {\rm \ km \ s^{-1} \ Mpc^{-1}}$,
\begin{equation}
{\rm SFR} \approx 32 \times
        \left( \frac{24}{f_{\rm lens}} \right) \times
        \left( \frac{f_{\rm dust}}{11} \right) \times
        \left( \frac{7.8 \times 10^{27} {\rm \ erg \ s^{-1} \ Hz^{-1}}}
                         {F_{\nu,1500}} \right) \times
        \left( \frac{2.5}{f_{\rm IMF}} \right) h^{-1} \ M_\sun \ {\rm yr}^{-1} \ , \label{eqn_SFR}
\end{equation}
where $f_{\rm lens}$ is the lensing magnification corresponding to the 
flux inside the annulus aperture, $f_{\rm dust}$ is the extinction at
rest wavelength 1500~\AA, $F_{\nu,1500}$ is the flux at rest 1500~\AA \
for a model forming stars at 1 $M_\sun {\rm \ yr^{-1}}$, and
$f_{\rm IMF}$ is a correction factor to the Salpeter IMF,
as described by \citet{pettini00} and references therein.
The numerical values given in eqn.~(\ref{eqn_SFR}) are for our fiducial
model with an age of 100 Myr and $E(B-V) = 0.25$. 
The corresponding numbers for the 10 Myr, $E(B-V) = 0.3$ 
and 1 Gyr, $E(B-V) = 0.2$ models are, respectively: 
${\rm SFR} = 18, 83 \ h^{-1} \ M_\sun \ {\rm yr}^{-1}$; 
$f_{\rm dust} = 17, 6.7$; and
$F_{\nu,1500} = 4.9, 8.6 \times 10^{27} {\rm \ erg \ s^{-1} \ Hz^{-1}}$.

As shown in Fig.~\ref{SED_models}, although an example dust-free 100 Myr 
model is clearly ruled out, all three of the dusty models considered 
give good matches to the data and cannot be distinguished given our
data, which cover just rest-frame UV wavelengths.  We defer a fuller
model-fitting analysis to a future paper that will also include 
the imaging and spectroscopic data we have obtained at observed-frame IR = 
rest-frame optical wavelengths, which will allow us to set much more
stringent constraints on model ages, dust extinction and star formation
rates \citep[e.g.,][]{ellingson96}.  Nevertheless, we can still compare
star formation rates obtained under similar assumptions for other systems.
In particular, for the fiducial 100 Myr model, the resulting SFR of
$32 \ h^{-1} \ M_\sun \ {\rm yr}^{-1}$ is about twice the value
of $17 \ h^{-1} \ M_\sun \ {\rm yr}^{-1}$ obtained by 
\citet{pettini00} for cB58 (after converting to our adopted cosmology),
but it is much lower than the $160 \ h^{-1} \ M_\sun \ {\rm yr}^{-1}$ obtained by 
\citet{allam07} for the 8 o'clock arc.  We can also compare our source
galaxy with the sample of about 100 (unlensed) $z \sim 2$
star-forming galaxies of \citet{erb06a,erb06b}, who derived SFRs
using detailed SED fitting with \citet{bc03} models, including 
additional IR data in the fits.  The bulk of the \citet{erb06a,erb06b}
sample galaxies were best fit with constant SFR models like the ones 
we have used.  Overall those galaxies have a mean best-fit age
of 1046 Myr, $E(B-V)$ of 0.15 and SFR of 
$52 \ M_\sun \ {\rm yr}^{-1}$ ($H_0 = 70 {\rm \ km \ s^{-1} \ Mpc^{-1}}$),
which becomes $26 \ h^{-1} M_\sun \ {\rm yr}^{-1}$ after converting
to our conventions.\footnote{We account for $H_0$, 
for a factor of 1.8 to convert the SFR normalization from the
\citet{chabrier03} IMF used by \citet{erb06a,erb06b} back to a 
Salpeter IMF, and finally for our use of 
the factor $f_{\rm IMF} = 2.5$ from \citet{pettini00}.}
We see that our source galaxy appears somewhat dustier, but otherwise 
it has a star formation rate close to the typical $z \sim 2$ star-forming 
galaxy from the \citet{erb06a,erb06b} sample.

\section{Conclusions}\label{sec:conclusions}

We have  reported on the discovery of the Clone system, consisting of
a star-forming, BX/BM-type Lyman break galaxy in the SDSS at a redshift 
of $z=2.001$, which is strongly lensed by a foreground luminous red galaxy 
at a redshift of $z=0.422$.  The lensed galaxy is remarkably bright, 
and at $r = 19.9$ it is among the brightest known lensed source galaxies 
with $z \geq 2$.

A simple SIE  lens model for the system  yields an  Einstein radius of
$\theta_{\rm   Ein}=3.82\pm0.03\arcsec$  (or $R_{\rm  Ein}=
14.8\pm0.1$\ $h^{-1}$~kpc at the lens redshift), a total  lensing  mass 
within the  Einstein radius of $2.10\pm0.03\times10^{12}\  h^{-1} M_{\sun}$,   
and a magnification factor for the lensed LBG of $27 \pm 1$. 
Combining the lens model with our follow-up $gVriz$ photometry, 
we have also estimated the (unlensed) star formation rate (SFR) of the 
source galaxy to be $32 \ h^{-1} \ M_\sun \ {\rm yr}^{-1}$, adopting a 
fiducial constant-SFR galaxy evolution model with an age of 
100 Myr and $E(B-V) = 0.25$.  Such a star formation rate is similar
to that found for samples of similar, but unlensed, $z \sim 2$ BX/BM galaxies.

We are pursuing a number of further follow-up observations on this system,
and we currently have optical and infrared data from HST Cycle 16, 
Spitzer Cycle 4, and Gemini North programs.  
The HST images indicate that the system
is more complex than can be seen from the ground, and 
analysis of that higher resolution data will help us investigate
the issues of substructure and image flux anomalies that we have
encountered in the lens modeling described here.  Moreover, we will
also use the SED information provided by the additional near-IR imaging 
and spectroscopy we are analyzing to better constrain the star formation
history and dust content than we have been able to do here using 
just the optical data.  These more detailed analyses will be the subjects 
of future papers that will exploit the rich follow-up data set that can 
be derived from this very bright high-redshift lensing system.



\acknowledgments

SSA acknowledges support from a HST Grant.  Support for program \#11167 was 
provided by NASA through a grant from the Space Telescope Science Institute, 
which is operated by the Association of Universities for Research in 
Astronomy, Inc., under NASA contract NAS5-26555.

This work was supported in part by Department of Energy contract
DE-AC02-76SF00515.

These results are based on observations obtained with the Apache Point
Observatory 3.5-meter telescope, which  is owned  and operated by  the
Astrophysical Research Consortium.

Based [in part] on data collected at Subaru Telescope,
which is operated by the National Astronomical Observatory of Japan.

Funding  for the SDSS  and  SDSS-II has been  provided  by  the Alfred
P.  Sloan Foundation,   the  Participating Institutions, the  National
Science   Foundation,  the  U.S. Department   of  Energy, the National
Aeronautics and Space Administration, the Japanese Monbukagakusho, the
Max Planck  Society, and  the  Higher  Education Funding  Council  for
England. The SDSS Web Site is http://www.sdss.org/.
The SDSS is managed by the Astrophysical Research Consortium for the 
Participating Institutions. The Participating Institutions are the 
American Museum of Natural History, Astrophysical Institute Potsdam, 
University of Basel, University of Cambridge, Case Western Reserve 
University, University of Chicago, Drexel University, Fermilab, the 
Institute for Advanced Study, the Japan Participation Group, Johns 
Hopkins University, the Joint Institute for Nuclear Astrophysics, 
the Kavli Institute for Particle Astrophysics and Cosmology, the 
Korean Scientist Group, the Chinese Academy of Sciences (LAMOST), 
Los Alamos National Laboratory, the Max-Planck-Institute for 
Astronomy (MPIA), the Max-Planck-Institute for Astrophysics (MPA), 
New Mexico State University, Ohio State University, University of 
Pittsburgh, University of Portsmouth, Princeton University, the 
United States Naval Observatory, and the University of Washington.

\clearpage



\clearpage


\clearpage

\clearpage

\clearpage

\clearpage

\clearpage

\clearpage

\clearpage

\clearpage

\begin{deluxetable}{ccccl}
\tablewidth{0pt}
\tablecaption{Observation Log\label{table_obslog}}
\tablehead{
\colhead{Filter/Grating} & 
\colhead{UT Date}     &
\colhead{Exposure}    & 
\colhead{Seeing}   &
\colhead{Notes}}
\startdata
\sidehead{Subaru 8.2m/FOCAS imaging}
 $V$ & 23 Jan 2007 & 3$\times 15$ sec & $0.53\arcsec$ & \\
\sidehead{Subaru 8.2m/FOCAS spectroscopy}
 300B+L600 & 23 Jan 2007 & 1$\times 600$ sec & $0.5\arcsec$ & slit includes knots A2, A3 \\
\sidehead{APO 3.5m/SPIcam imaging}
 $g$ & 11 Jan 2008 & 3$\times 300$ sec & $1.0\arcsec$ & \\
 $r$ & 11 Jan 2008 & 3$\times 300$ sec & $1.2\arcsec$ & \\
 $i$ & 11 Jan 2008 & 3$\times 300$ sec & $0.9\arcsec$ & \\
 $z$ & 28 Oct 2007 & 3$\times 300$ sec & $1.1\arcsec$ & \\
\sidehead{APO 3.5m/DIS spectroscopy}
B400/R300 & 19 Nov 2007 & 2$\times 600$ sec & $\sim$1.5$\arcsec$ & slit includes knots A1, A2 \\
\enddata
\end{deluxetable}

\clearpage

\begin{deluxetable}{lcccc}
\tabletypesize{\scriptsize}
\tablewidth{0pt}
\tablecaption{Arc Spectroscopic Features\label{table_spectra}}
\tablehead{
\colhead{} & 
\multicolumn{2}{c}{Subaru 8.2m (A2+A3)} &
\multicolumn{2}{c}{APO 3.5m (A1+A2)} \\
\colhead{ID/Rest Wavelength} &
\colhead{Observed Wavelength\tablenotemark{a}} & 
\colhead{Redshift} & 
\colhead{Observed Wavelength\tablenotemark{a}} & 
\colhead{Redshift} \\
\colhead{[\AA]} &
\colhead{[\AA]} &
\colhead{} &
\colhead{[\AA]} &
\colhead{}
}
\startdata
Ly$\alpha$ 1215.7      &   \nodata & \nodata            &    3650.0\tablenotemark{b} &  2.0025 \\
SiII 1260.4            &   \nodata & \nodata            &    3781.4 &  2.0002 \\
OI 1302.2, SiII 1304.4  &   \nodata & \nodata            &    3904.6 &  1.9959 \\
CII 1334.5             &    3999.1 &  1.9967            &    4002.6 &  1.9993 \\
SiIV 1393.8            &    4180.3 &  1.9992            &    4181.8 &  2.0003 \\
SiIV 1402.8            &    4210.4 &  2.0014            &    4209.1 &  2.0005 \\
SiII 1526.7            &    4584.0 &  2.0026            &    4581.9 &  2.0012 \\
CIV 1548.2,1550.8      &    4650.2 &  2.0011            &    4644.7 &  1.9975 \\
FeII 1608.4            &    4831.7 &  2.0040            &    4827.4 &  2.0014 \\
AlII 1670.8            &    5015.6 &  2.0019            &    5015.2 &  2.0017 \\
\hline
mean redshift          &           &  2.0010$\pm$0.0009 &           &  2.0001$\pm$0.0006 \\
\hline
\enddata
\tablenotetext{a} {\ The observed wavelengths were converted from wavelengths in air to wavelengths in vacuum 
using eq. [3] of \citet{morton91}.}
\tablenotetext{b} {\ Ly$\alpha$ is seen in absorption in the APO 3.5m 
spectrum, with an observed equivalent width of $-15 \pm 1$~\AA, 
or $-5 \pm 0.3$~\AA \ in the rest frame.}
\end{deluxetable}

\clearpage

\begin{deluxetable}{lcccccc}
\rotate
\tabletypesize{\scriptsize}
\tablewidth{0pt}
\tablecaption{GALFIT Modeling Results\label{galfit_table}}
\tablehead{
\colhead{Object} & 
\colhead{Model} &
\colhead{$V$ Magnitude\tablenotemark{a}}    &
\colhead{Effective Radius $r_e$ (arcsec)}    &
\colhead{Exponent}    &
\colhead{Axis Ratio}  & 
\colhead{Position Angle (deg E of N)}} 
\startdata
B(LRG)&Sersic&$18.97\pm 0.05$   &$3.67\pm0.34$  &$4.68\pm0.22$ &$0.89\pm 0.01$ &$-68.2\pm6.2$\\
C&de Vaucouleurs&$22.54\pm 0.17$   &$0.56\pm0.16$ &$4$ &$0.59\pm 0.15$ &$47.3\pm12.5$\\
D&de Vaucouleurs&$22.84\pm 0.09$   &$0.19\pm0.1$ &$4$ &$0.77\pm 0.2$
&$42.7\pm34.6$\\
\tableline
& & & Scalelength $r_s$ (arcsec) & & & \\ 
\tableline
Arc (A3)&Exponential &$20.88\pm 0.01$ &$0.52\pm0.01$ & &$0.24\pm 0.01$ &$68.4\pm0.5$\\
Arc &Exponential &$21.72\pm 0.04$ &$0.79\pm0.04$ & &$0.05\pm 0.01$ &$39.4\pm0.66$\\
Arc (A2)&Exponential &$21.64\pm 0.07$ &$0.47\pm0.02$ & &$0.15\pm 0.02$ &$12.6\pm1.2$\\
Arc (A2)&Exponential &$21.82\pm 0.08$ &$0.48\pm0.03$ & &$0.03\pm 0.0$ &$6.5\pm0.6$\\
Arc (A1)&Exponential &$20.91\pm 0.01$ &$0.57\pm0.01$ & &$0.22\pm 0.01$ &$-28.0\pm0.4$\\
Counter-image (A4)&Exponential &$22.43\pm 0.03$ & $0.23\pm0.01$& &$0.28\pm 0.06$ &$25.8\pm3.2$\\
\enddata
\tablenotetext{a}{The photometry errors given here are the formal errors
reported by GALFIT.  The magnitudes have been corrected for Milky Way
extinction, using values from the SDSS DR5 database, which are
in turn based on the dust maps of \citet{schlegel98}.
Specifically, the extinction correction in $V$ is
0.070 mag and $E(B-V) = 0.023$.}
\end{deluxetable}

\clearpage

\begin{deluxetable}{lrrccccc}
\tabletypesize{\scriptsize}
\tablewidth{0pt}
\tablecaption{Photometry Results\label{table_photometry}\tablenotemark{b,c}}
\tablehead{
\colhead{Object\tablenotemark{a}} & 
\colhead{RA (deg)} &
\colhead{Dec (deg)} &
\colhead{$g$} &
\colhead{$V$} &
\colhead{$r$} &
\colhead{$i$} &
\colhead{$z$}}
\startdata
B (LRG, GALFIT) & 181.508732 & 51.708196 & $20.02 \pm 0.02$ & $18.97 \pm 0.06$ & $18.24 \pm 0.02$ & $17.59 \pm 0.02$ & $17.17 \pm 0.03$  \\
lens light aperture & & & $ 20.66 \pm   0.03$ & $ 19.62 \pm   0.04$ & $ 18.92 \pm   0.02$ & $ 18.27 \pm   0.02$ & $ 17.85 \pm   0.04$ \\
\tableline
C (GALFIT) & 181.508354 & 51.708526 & $23.33 \pm 0.08$ & $22.54 \pm 0.17$ & $22.24 \pm 0.08$ & $21.59 \pm 0.05$ & $21.11 \pm 0.15$  \\
D (GALFIT) & 181.509529 & 51.708270 & $23.90 \pm 0.09$ & $22.84 \pm 0.09$ & $22.27 \pm 0.04$ & $21.70 \pm 0.04$ & $21.22 \pm 0.10$  \\
E (fit)   & & & $ 24.29 \pm   0.13$ & $ 23.24 \pm   0.13$ & $ 22.51 \pm   0.13$ & $ 21.86 \pm   0.13$ & $ 21.44 \pm   0.13$ \\
\tableline
A1 (3$\arcsec$) & 181.507131 & 51.707665 & $ 21.61 \pm   0.03$ & $ 21.26 \pm   0.04$ & $ 21.04 \pm   0.03$ & $ 20.67 \pm   0.03$ & $ 20.56 \pm   0.09$ \\
A1 (decon, 3$\arcsec$)  & & & $ 21.71 \pm   0.04$ & $ 21.45 \pm   0.05$ & $ 21.36 \pm   0.06$ & $ 21.11 \pm   0.08$ & $ 21.20 \pm   0.19$ \\
\tableline
A2 (3$\arcsec$) & 181.506902 & 51.708333 & $ 21.67 \pm   0.03$ & $ 21.43 \pm   0.04$ & $ 21.28 \pm   0.03$ & $ 20.98 \pm   0.03$ & $ 20.86 \pm   0.11$ \\
A2 (decon, 3$\arcsec$)  & & & $ 21.71 \pm   0.03$ & $ 21.52 \pm   0.05$ & $ 21.44 \pm   0.05$ & $ 21.21 \pm   0.07$ & $ 21.18 \pm   0.17$ \\
\tableline
A3 (3$\arcsec$) & 181.508016 & 51.709280 & $ 21.39 \pm   0.03$ & $ 21.14 \pm   0.04$ & $ 21.09 \pm   0.03$ & $ 20.86 \pm   0.03$ & $ 20.84 \pm   0.11$ \\
\tableline
annulus    & & & $ 20.22 \pm   0.02$ & $ 19.96 \pm   0.04$ & $ 19.82 \pm   0.02$ & $ 19.53 \pm   0.02$ & $ 19.50 \pm   0.07$ \\
annulus (decon) & & & $ 20.25 \pm   0.02$ & $ 20.02 \pm   0.05$ & $ 19.93 \pm   0.04$ & $ 19.68 \pm   0.06$ & $ 19.73 \pm   0.12$ \\
\tableline
A4 (GALFIT) & 181.510019 & 51.707975 & $22.77 \pm 0.04$ & $22.43 \pm 0.04$ & $22.42 \pm 0.04$ & $22.24 \pm 0.05$ & $22.23 \pm 0.22$  \\
\enddata
\tablenotetext{a}{The type of photometry measurement and/or
the photometry aperture is indicated in the parentheses next to the 
object name.  Detailed descriptions of the photometry measurements
and techniques are given in \S\ref{subaru_photometry}, \S\ref{lens_modeling},
and \S\ref{apo_photometry}.}
\tablenotetext{b}{The photometry errors are taken to be the formal 
GALFIT errors, or the statistical errors from photon noise in the 
photometry aperture, added in quadrature to the rms photometric calibration
errors for the SDSS DR5 \citep{dr5}.}
\tablenotetext{c}{The magnitudes have been corrected for Milky Way
extinction, using values from the SDSS DR5 database, which are
in turn based on the dust maps of \citet{schlegel98}.
Specifically, the extinction corrections in $gVriz$ are 
0.086, 0.070, 0.063, 0.047, and 0.034 mag, respectively, 
and $E(B-V) = 0.023$.}
\end{deluxetable}





\begin{thebibliography}{}
\bibitem[Adelman-McCarthy et al.(2006)]{dr4} 
   Adelman-McCarthy, J., et al.\ 2006, \apjs, 162, 38
\bibitem[Adelman-McCarthy et al.(2007)]{dr5}
   Adelman-McCarthy, J., et al.\ 2007, \apjs, 172, 634
\bibitem[Allam et al.(2007)]{allam07} 
   Allam, S., et al.\ 2007, \apj, 662, 51
\bibitem[Allam et al.(2008)]{hstpaper}
   Allam, S., et al.\ 2008, in preparation
\bibitem[Bahcall et al.(1995)]{bahcall95}
   Bahcall, N.~A., Lubin, L.~M., \& Dorman, V.\ 1995, \apjl, 447, L81
\bibitem[Belokurov et al.(2007)]{belokurov07}
   Belokurov, V., et al.\ 2007, \apjl, 671, L9
\bibitem[Bertin \& Arnouts(1996)]{bertin96}
   Bertin, E., \& Arnouts, S.\ 1996, \aaps, 117, 393
\bibitem[Bolton et al.(2006)]{bolton06}
   Bolton, A.~S., et al.\ 2006, \apj, 638, 703
\bibitem[Bruzual \& Charlot(2003)]{bc03}
   Bruzual, G., \& Charlot, S.\ 2003, \mnras, 344, 1000
\bibitem[Calzetti et al.(2000)]{calzetti00}
   Calzetti, D., Armus, L., Bohlin, R.~C., Kinney, A.~L.,
   Koornneef, J., \& Storchi-Bergmann, T.\ 2000, \apj, 533, 682
\bibitem[Chabrier(2003)]{chabrier03}
   Chabrier, G.\ 2003, \pasp, 115, 763
\bibitem[Coleman et al.(1980)]{coleman80}
   Coleman, G.~D., Wu, C.-C., \& Weedman, D.~W.\ 1980, \apjs, 43, 393
\bibitem[Ellingson et al.(1996)]{ellingson96}
   Ellingson, E., Yee, H.~K.~C., Bechtold, J., \& Elston, R. 1996,
   \apjl, 466, L71
\bibitem[Erb et al.(2006a)]{erb06a}
   Erb, D.~K., Steidel, C.~C., Shapley, A.~E., Pettini, M., 
   Reddy, N.~A., \& Adelberger, K.~L. 2006a, \apj, 646, 107
\bibitem[Erb et al.(2006b)]{erb06b}
   Erb, D.~K., Steidel, C.~C., Shapley, A.~E., Pettini, M., 
   Reddy, N.~A., \& Adelberger, K.~L. 2006b, \apj, 647, 128
\bibitem[Fukugita et al.(1996)]{fukugita96}
   Fukugita, M., Ichikawa,  T., Gunn, J.~E., Doi, M., Shimasaku, K., 
   \& Schneider, D.~P.\ 1996, \aj, 111, 1748
\bibitem[Hansen et al.(2005)]{hansen05}
   Hansen, S.~M., et al.\ 2005, \apj, 633, 122
\bibitem[Jester et al.(2005)]{jester05} 
   Jester, S., et al.\ 2005, \aj, 130, 873
\bibitem[Kashikawa et al.(2002)]{kashikawa02} 
   Kashikawa, N., et al.\ 2002, \pasj, 54, 819
\bibitem[Keeton et al.(2001)]{keeton01}
   Keeton, C.~R.\ 2001, arXiv:astro-ph/0102340
\bibitem[Keeton et al.(2005)]{keeton05}
   Keeton, C.~R., Gaudi B.~S., \& Petters, A.O.\ 2005, \apj, 635, 35
\bibitem[Koester et al.(2007)]{koester07}
   Koester, B.~P., et al.\ 2007, \apj, 660, 221
\bibitem[Koopmans et al.(2006)]{koopmans06}
   Koopmans, L.~V.~E., et al.\ 2006, \apj, 649, 599
\bibitem[Kormann et al.(1994)]{kormann94} 
   Kormann, R., Schneider, P., \& Bartelmann, M.\ 1994, \aap, 284, 285
\bibitem[Kubik(2007)]{kubik07}
   Kubik, D.\ 2007, M.S. Thesis, Northern Illinois University
\bibitem[Morton(1991)]{morton91} 
   Morton, D.~C.\ 1991, \apjs, 77, 119
\bibitem[Navarro et al.(1997)]{navarro97}
   Navarro, J.~F., Frenk, C.~S., \& White, S.~D.~M.\ 1997, \apj, 490, 493
\bibitem[Oguri(2006)]{oguri06} 
   Oguri, M.\ 2006, \mnras, 367, 1241
\bibitem[Peng et al.(2002)]{peng02} 
   Peng, C., et al.\ 2002, \aj, 124, 266
\bibitem[Pettini et al.(2000)]{pettini00} 
   Pettini, M., Steidel, C.~C., Adelberger, K.~L.,
   Dickinson, M., \& Giavalisco, M.\ 2000, \apj, 528, 96
\bibitem[Salpeter(1955)]{salpeter55}
   Salpeter, E.~E.\ 1955, \apj, 121, 161
\bibitem[Schlegel et al.(1998)]{schlegel98}
   Schlegel, D.~J., Finkbeiner, D.~P., \& Davis, M.\ 1998, \apj, 500, 525
\bibitem[Shapley et al.(2003)]{shapley03}
   Shapley, A.~E., et al.\ 2003, \apj, 588, 65
\bibitem[Smail et al.(2007)]{smail07}
   Smail, I., et al.\ 2007, \apjl, 654, L33
\bibitem[Steidel et al.(2003)]{steidel03}
   Steidel, C.~C., et al.\ 2003, \apj, 592, 728
\bibitem[Steidel et al.(2004)]{steidel04}
   Steidel, C.~C., Shapley, A.~E., Pettini, M., Adelberger, K.~L.,
   Erb, D.~K., Reddy, N.~A., \& Hunt, M.~P.\ 2004, \apj, 604, 534
\bibitem[Teplitz et al.(2000)]{teplitz00}
   Teplitz, H.~I., et al.\ 2000, \apjl, 533, L65
\bibitem[Treu et al.(2006)]{treu06}
   Treu, T., et al.\ 2006, \apj, 640, 662
\bibitem[Wallington et al.(1996)]{wallington96} 
   Wallington, S., et al.\ 1996, \apj, 465 64
\bibitem[Wayth \& Webster(2006)]{wayth06} 
   Wayth, R., \& Webster, R.L.\ 2006, \mnras, 1187, 372  
\bibitem[Yee et al.(1996)]{yee96}
   Yee, H.~K.~C., et al.\ 1996, \aj, 111, 1783
\bibitem[York et al.(2000)]{york00}
   York, D.~G., et al.\ 2000, \aj, 120, 1579

\end{thebibliography}
\end{document}